\documentclass{aastex}
\usepackage{amssymb,amsmath,natbib,gensymb}
\usepackage{graphicx,graphics}
\usepackage{multirow}
\usepackage{subfigure}
\usepackage[colorlinks,linkcolor=blue, anchorcolor=blue, citecolor=blue,filecolor=blue,menucolor=blue,runcolor=blue,urlcolor=blue]{hyperref}
\usepackage{geometry}
\newcommand{\cii}{[C\,{\sc ii}]}

\begin{document}

\title{Star Formation and ISM Properties in Host Galaxies of Three Far-infrared Luminous Quasars at $z\sim6$}

\author{Yali Shao\altaffilmark{1, 2}$^{,\dagger}$, Ran Wang\altaffilmark{2}$^{,\bigstar}$, Chris L. Carilli\altaffilmark{3, 5}, Jeff Wagg\altaffilmark{4}, Fabian Walter\altaffilmark{5, 6, 13}, Jianan Li\altaffilmark{1, 2}, Xiaohui Fan\altaffilmark{7}, Linhua Jiang\altaffilmark{2}, Dominik A. Riechers\altaffilmark{8}, Frank Bertoldi\altaffilmark{9}, Michael A. Strauss\altaffilmark{10},  Pierre Cox\altaffilmark{11}, Alain Omont\altaffilmark{11}, Karl M. Menten\altaffilmark{12}}

\altaffiltext{1}{Department of Astronomy, School of Physics, Peking University, Beijing 100871, China}
\altaffiltext{2}{Kavli Institute for Astronomy and Astrophysics, Peking University, Beijing 100871, China}
\altaffiltext{3}{Cavendish Laboratory, 19 J. J. Thomson Avenue, Cambridge CB3 0HE, UK}
\altaffiltext{4}{SKA Organization, Lower Withington Macclesfield, Cheshire SK11 9DL, UK}
\altaffiltext{5}{National Radio Astronomy Observatory, Socorro, NM 87801-0387, USA}
\altaffiltext{6}{Max-Planck-Institut for Astronomie, K\"{o}nigstuhl 17, D-69117 Heidelberg, Germany}
\altaffiltext{7}{Steward Observatory, University of Arizona, 933 North Cherry Avenue, Tucson, AZ 85721, USA}
\altaffiltext{8}{Department of Astronomy, Cornell University, 220 Space Sciences Building, Ithaca, NY 14853, USA}
\altaffiltext{9}{Argelander-Institut f\"{u}r Astronomie, University at Bonn, Auf dem H\"{u}gel 71, D-53121 Bonn, Germany}
\altaffiltext{10}{Department of Astrophysical Sciences, Princeton University, Princeton, NJ 08544, USA}
\altaffiltext{11}{Institut d'Astrophysique de Paris, Sorbonne Universit\'{e}, CNRS, UMR 7095, 98 bis bd Arago, 75014 Paris, France}
\altaffiltext{12}{Max-Planck-Institut fur Radioastronomie, Auf dem H\"{u}gel 69, 53121 Bonn, Germany}
\altaffiltext{13}{Astronomy Department, California Institute of Technology, MC105-24, Pasadena, CA 91125, USA}
\altaffiltext{$\dagger$}{E-mail: \href{mailto:shaoyali0922@pku.edu.cn}{shaoyali0922@pku.edu.cn}}
\altaffiltext{$\bigstar$}{E-mail: \href{mailto:rwangkiaa@pku.edu.cn}{rwangkiaa@pku.edu.cn}}

\begin{abstract}
We present Karl G. Jansky Very Large Array (VLA) observations of the CO (2$-$1) line emission towards three far-infrared luminous quasars at $z\sim6$: SDSS J231038.88$+$185519.7 and SDSS J012958.51$-$003539.7 with  $\sim0\farcs6$ resolution and SDSS J205406.42$-$000514.8 with $\sim2\farcs1$ resolution. All three sources are detected in the CO (2$-$1) line emission -- one source is marginally resolved, and the other two appear as point sources. Measurements of the CO (2$-$1) line emission allow us to calculate the molecular gas mass even without a CO excitation model. The inferred molecular gas masses are (0.8$-$4.3) $\times$ 10$^{10}$ $M_{\odot}$. The widths and redshifts derived from the CO (2$-$1) line are consistent with previous CO (6$-$5) and [\ion{C}{2}] measurements. We also report continuum measurements using the Herschel for  SDSS J231038.88$+$185519.7 and SDSS J012958.51$-$003539.7, and for SDSS J231038.88+185519.7, data obtained at $\sim140$ and $\sim300$ GHz using the Atacama Large Millimeter/submillimeter Array (ALMA).  In the case of SDSS J231038.88+185519.7, we present a detailed analysis of the spectral energy distribution and derive the dust temperature ($\sim40$ K), the dust mass ($\sim10^{9}$ $M_{\odot}$), the far-infrared luminosity (8$-$1000 $\mu$m; $\sim10^{13}$ $ L_{\odot}$) and the star formation rate (2400$-$2700 $M_{\odot}$ yr$^{-1}$). Finally, an analysis of the photo-dissociation regions associated with the three high redshift quasars indicates that the interstellar medium in these sources has similar properties to local starburst galaxies.   

\end{abstract}

\keywords{galaxies: evolution --- galaxies: active --- galaxies: high redshift --- submillimeter: galaxies --- quasars: general --- radio line: galaxies}

\section{Introduction}

More than 200 quasars have been discovered above redshift 5.7 (e.g., \citealt{Fan2006lala}; \citealt{Mortlock2009, Mortlock2011}; \citealt{Willott2009, Willott2010eddington,Willott2010new};  \citealt{Venemans2013, Venemans2015a}; \citealt{Banados2014, Banados2016, Banados2018}; \citealt{Jiang2015, Jiang2016};  \citealt{Matsuoka2016, Matsuoka2018}; \citealt{Mazzucchelli2017}). These high redshift quasars are key to understand the co-evolution between supermassive black holes (SMBHs) and their host galaxies at the end of the reionization epoch. Observations of the dust, molecular and atomic gas content of these objects allow us to probe their star formation activity and derive their interstellar medium properties.

The rest frame far-infrared (FIR) continuum emission in these sources originates mainly from dust heated by ultraviolet (UV) radiation from young and massive stars in the host galaxies. 
Submillimeter and millimeter observations using, e.g., the instruments on Herschel, the Max Plank Millimeter Bolometer Array (MAMBO) on the IRAM-30 m telescope or the Submillimeter Common User Bolometer Array (SCUBA) on the James Clerk Maxwell Telescope (JCMT) (\citealt{Bertoldi2003}; \citealt{Petric2003}; \citealt{Robson2004}; \citealt{Beelen2006}; \citealt{Wang2007, Wang2008SHARC, Wang2008thermal, Wang2011fir}; \citealt{Leipski2013, Leipski2014}), and the ALMA (\citealt{Decarli2018}; \citealt{Venemans2018}), have detected dust continuum in the host galaxies of many $z\sim6$ quasars, with FIR luminosities of $\sim10^{11-13}$ $L_{\odot}$, and dust masses on the order of 10$^{7-9}$ $M_{\odot}$. The most luminous objects have FIR luminosities similar to those of the ultraluminous infrared galaxies (ULIRGs; $L_{\rm FIR} > 10^{12} L_{\odot}$) and hyper-luminous infrared galaxies (HLIRGs; $L_{\rm FIR} > 10^{13} L_{\odot}$) in the local universe (e.g., \citealt{Rosenberg2015}), indicating that they are forming stars with star formation rates (SFRs) of a  few tens to thousands $M_{\odot}$ yr$^{-1}$, co-eval with rapid SMBH accretion.

Most of these high-$z$ FIR-luminous quasars have been detected in carbon monoxide (CO). As principal molecular tracer, CO provides a tool to probe the physical conditions of star-forming gaseous reservoirs, through multiple transitions redshifted into the submillimeter/millimeter range. 

Intermediate to high-$J$ transitions of CO from 3$-$2 to 9$-$8 have been detected with the VLA, the IRAM Plateau de Bure interferometer and its successor - the Northern Extended Millimeter array (NOEMA) and the ALMA in many high redshift quasars (\citealt{Bertoldi2003}; \citealt{Walter2003, Walter2004}; \citealt{Carilli2007}; \citealt{Riechers2009}; \citealt{Wang2010, Wang2011fir, Wang2013, WangRan2016}; \citealt{Venemans2017a, Venemans2017b}; \citealt{Feruglio2018}; Li et al. in preparation). \citet{Walter2004} observed the CO (3$-$2)  line emission from SDSS J114816.64+525150.3 (hereafter 1148+5251) at $z = 6.42$ using the VLA, and measured a CO source size of 3.6 kpc  $\times$  1.4 kpc (full width at half maximum; FWHM). They also derived a dynamical mass of 4.5  $\times$  10$^{10}$ $M_{\odot}$, which is less than the  stellar bulge mass (of order 10$^{12}$ $M_{\odot}$) predicted by the present-day $M_{\rm BH}-M_{\rm bulge}$ relation \citep{Ho2013}, which may indicate a faster SMBHs evolution than their hosts. 

Observations of the low-$J$ CO transitions ($J_{\rm upper}$ $\leqslant$ 2) in $z\sim6$ quasars are very difficult due to the low flux density and limited telescope sensitivity. Only eight $z\sim6$ quasars have been observed in CO (2$-$1) line emission ($\nu_{\rm rest}$ = 230.538 GHz) all using the VLA Ka band, and five of them have been detected (\citealt{Wang2011, WangRan2016}; \citealt{Stefan2015}; \citealt{Venemans2017b}). The VLA is the only instrument that can observe the CO (2$-$1) line with proper frequency coverage and high sensitivity for $z\sim6$ objects. The CO (2$-$1) line emission allows us to measure the molecular gas mass directly with a molecular gas mass conversion factor. The CO (2$-$1) line is also a crucial tracer to probe the low-$J$ part of CO spectral line energy distributions (SLEDs),  and the excitation, the spatial distribution and the surface density of the extended cold gas in star-forming quasar host galaxies. For example, by observing the $z\sim5.7$ quasar SDSS J092721.82+200123.7,  \citet{Wang2011} constrained the CO (2$-$1) line excitation in the central $\sim10$ kpc of the source, and  estimated a molecular gas mass of order of 10$^{10}$ $M_{\odot}$. The redshift, the line width and the gas mass derived from CO (2$-$1) in this quasar are consistent with those from CO (6$-$5) and CO (5$-$4) observations in \citet{Carilli2007}. However, the CO (2$-$1) line has been detected only toward a few $\sim6$ quasars. In this paper we increase the sample with CO (2$-$1) detections by a factor of 60$\%$. These data will be critical in investigations of the CO SLEDs with multiple CO transition observations in the future.

In this paper, we report VLA observations of CO (2$-$1) line emission in three FIR-luminous high redshift quasars: SDSS J231038.88$+$185519.7 (hereafter J2310$+$1855 at $z = $ 6.0), SDSS J012958.51$-$003539.7 (hereafter J0129$-$0035 at $z = $ 5.7), and SDSS J205406.42$-$000514.8 (hereafter J2054$-$0005 at $z = $ 6.0). We complement these data with continuum measurements from Herschel Photodetector Array Camera and Spectrometer instrument (PACS; \citealt{Poglitsch2010}) and Spectral and Photometric Imaging Receiver (SPIRE; \citealt{Griffin2010}) for J2310$+$1855 and J0129$-$0035, and in the case of J2310+1855, with ALMA. This paper is organized as follows. In Section \ref{obs} we describe the sample, the observations and the data reduction. In Section \ref{red} we present our results. In Section \ref{dis}, we first discuss the gas distribution and the gas mass of the three targets; then we analyze the dust temperature, the dust mass and the star formation rate of J2310$+$1855 through a fit to the continuum spectral energy distribution (SED); finally we discuss models for PDRs in the three targets. In Section \ref{sec5}, we present a short summary. Throughout this work we assume a $\Lambda$CDM cosmology with $H_{0}$ = 71 km s$^{-1}$ Mpc$^{-1}$, $\Omega_{\rm M}$ = 0.27 and $\Omega_{\Lambda}$ = 0.73 \citep{Spergel2007}.

\section{Observations and data reduction}
\label{obs}

We selected the three high redshift quasars - J2310$+$1855, J0129$-$0035 and J2054$-$0005 from our previous  CO (6$-$5) surveys (\citealt{Wang2010}; \citealt{Wang2011fir}). They are all FIR-luminous quasars and contain a lot of gas.
We carried out VLA observations of the CO (2$-$1) line emission in the three objects. FIR continuum data for J2310$+$1855 and J0129$-$0035 were also obtained from Herschel PACS and SPIRE, and measurements of the continuum at 140 and 300 GHz were made using the ALMA in the case of J2310+1855. We list the details and the observations of the three sources in Table \ref{tab1}.

\begin{deluxetable}{lcccccccccc}
\tabletypesize{\scriptsize} 
\tablecaption{Sample and observations\label{tab1}}
\tablewidth{0pt}
\tablecolumns{8}
\tablehead{\colhead{Source}&\multicolumn{5}{c}{Herschel}&\colhead{}&\multicolumn{2}{c}{VLA}&\colhead{}&\colhead{ALMA} \\
          \cline{2-6} \cline{8-9} \\
          \colhead{}& \multicolumn{2}{c}{Herschel PACS} &\colhead{}& \multicolumn{2}{c}{Herschel SPIRE}&\colhead{}&\multicolumn{2}{c}{}&\colhead{}&\colhead{}\\
          \cline{2-3} \cline{5-6}\\
          \colhead{} & \colhead{OD} & \colhead{OBSIDs}  &\colhead{} &\colhead{OD} & \colhead{OBSIDs} &\colhead{}& \colhead{$t_{\rm obs}$}& \colhead{Configuration}&\colhead{}& \colhead{$t_{\rm obs}$}\\
          \colhead{(1)} & \colhead{(2)} & \colhead{(3)} & \colhead{} & \colhead{(4)} & \colhead{(5)} &\colhead{} & \colhead{(6)} & \colhead{(7)}&\colhead{} & \colhead{(8)}}
\startdata
J2310$+$1855   &1122                        &13422468221/1342246822          &&1314            &1342257359         &&32 h   &C&&5.5 h\\
J0129$-$0035   &-                             &-                               &&1330            &1342258372         &&15 h   &C&&-\\
J2054$-$0005    &-                             &-                               &&-                 &-& &20 h   &D&&-
\enddata
\tablecomments{Column 1: source name; Column 2$-$5: operational day (OD) and the unique IDs of the Herschel PACS and SPIRE observations; Column 6$-$8: total exposure time and configuration of the VLA and ALMA observations. }
\end{deluxetable}

\subsection{VLA}
The CO (2$-$1) observations were performed using the Ka band receivers (centered at 32 GHz) of the VLA. The observations of J2310+1855 were carried out on 2012 January 28, and 2014 October 14 and 15 using the C-configuration. The total observing time was 15 hours and 8 hours on-source. For J0129$-$0035, the data were taken between 2013 July 8 and August 6 also in the C-configuration, with a total observing time of 32 hours and 17 hours on-source. Finally, J2054$-$005 was observed from 2014 June 29 to July 21 using the D-configuration, with a total observing time of 20 hours and 10.5 hours on-source. We used the WIDAR 8-bit samplers to maximize the line sensitivity. The total bandwidth is 2 GHz with 16 128 MHz-wide sub-bands in full polarization mode, and we centred the CO (2$-$1) line in one of the 128 MHz sub-bands. The reference redshifts of these sources were based on the previous CO (6$-$5) observations ($\nu_{\rm obs}$ = 32.922 GHz for J2310+1855; $\nu_{\rm obs}$ = 34.006 GHz for J0129$-$0035; $\nu_{\rm obs}$ = 32.757 GHz for J2054$-$005; \citealt{Wang2010, Wang2011}). Flux calibrations were performed with the standard VLA calibrators: 3C286 and 3C48. The spatial resolutions achieved during these observations are $0\farcs6$ for C-configuration and $2\farcs1$ for D-configuration.

The data were reduced using the Common Astronomy Software Application (CASA\footnote{\url{https://casa.nrao.edu/}}) standard pipeline. The final data cubes were reduced by the CLEAN task using a robust weighting factor of 0.5 to optimize the noise per frequency bin and the resolution of the final map.

\subsection{Herschel}
\subsubsection{PACS}
J2310$+$1855 was observed by Herschel PACS at 100 and 160 $\mu$m using the mini-scan map observing template. We executed the observations over two different scan angles with observing parameters as recommended in the mini-scan map Astronomical Observation Template (AOT) release note. Fourteen repetitions were employed for each scan direction. The final on-source integration time was 1792 seconds.

Data reduction was performed within the Herschel Interactive Processing Environment (HIPE; version 15.0.1; \citealt{Ott2010}). We followed the standard pipeline for PACS mini-scan observations. Each scan direction was processed individually and mosaicked at the end of the procedure.  Aperture photometry of the final mosaics was also performed using HIPE. We used aperture sizes of 6$\arcsec$ and 9$\arcsec$ radii in 100 and 160 $\mu$m bands, respectively. The residual sky emission was derived in a sky annulus between 20$\arcsec$ and 25$\arcsec$ (100 $\mu$m map) or 24$\arcsec$ and 28$\arcsec$ (160 $\mu$m map). Aperture corrections were determined from the encircled energy fraction of unresolved sources provided by calibrator observations. The photometric uncertainties cannot be measured directly from the pixel-to-pixel variations because the final PACS maps are heavily influenced by correlated noise. We followed \citet{Leipski2013, Leipski2014} to determine the photometric uncertainties: first we masked the centre source with the same aperture size used in quasar photometry, then randomly placed $\sim1000$ apertures on the images with the same diameter as used for photometry. In order to exclude the noisy edge of the maps, we restricted ourselves to the region of sky with 75$\%$ or more integration time compared with the position of the quasar. Then we fitted a Gaussian to the measured fluxes in 1000 apertures. We set the 1$\sigma$ photometric uncertainty to the sigma value of the Gaussian profile of the final map. The Herschel PACS photometric results for J2310+1855 are listed in Table \ref{conpho}.

\subsubsection{SPIRE}
Herschel SPIRE observations towards J2310$+$1855 and J0129$-$0035 were carried out at 250, 350, and 500 $\mu$m using small scan map mode with 9 repetitions for each object. The total on-source integration time per source was 370 seconds. SPIRE observations are dominated by the confusion noise (\citealt{Nguyen2010}; PSW: 5.8 mJy beam$^{-1}$; PMW: 6.3 mJy beam$^{-1}$; PLW: 6.8 mJy beam$^{-1}$).

Data reduction was executed in HIPE (version 15.0.1) following the standard SPIRE small scan observation pipeline. We used the HIPE built-in source extractor called `sourceExtractorSussextractor' (\citealt{Savage2007}) to determine the source location and the flux density. We estimated photometric uncertainties following the method in \citet{Leipski2013, Leipski2014}: first we created an artificial source image with all sources found by `sourceExtractorSussextractor' in the calibrated map, then subtracted it from the observed map to get the residual map. On this residual map we determined the pixel-to-pixel $rms$ in a reasonable box with size of $\sim8$ times the FWHM ($18\farcs2$, $24\farcs9$, and $36\farcs3$ for default map pixel sizes of 6$\arcsec$, 10$\arcsec$, and 14$\arcsec$ at 250, 350, and 500 $\mu$m, individually) and centered at the position of the quasar in order to have enough sampling of quasar surrounding sky and avoid lower coverage regions of noisy edged sky. For the J2310$+$1855 SPIRE observations in particular, the sky is not uniform due to bright foreground sources, so we calculated the flux noise as follows: first we added 100 fake sources with flux densities equal to that of the target with a Gaussian scatter of 3 mJy in the raw data, and used standard procedures to reduce the data to generate the scientific image. The fake sources are in source-free areas with good coverage (e.g. $>$60$\%$) and are near targets (e.g. within $\sim8$ times the FWHM with targets at the center). Then we used `sourceExtractorSussextractor' to estimate source flux density. If the source was not detected, we measured the flux density at the input position. The quoted photometric uncertainty was the $rms$ difference between the input and output fluxes. The final Herschel SPIRE photometric results for J2310+1855 and J0129$-$0035 are presented in Table \ref{conpho}.

\subsection{ALMA}
\label{alma}
We observed the dust continuum emission around 140 and 300 GHz towards J2310+1855 during ALMA Cycle 3 (program ID: 2015.1.01265.S; PI: Ran Wang). This program aims to observe CO (9$-$8), CO (8$-$7), [\ion{N}{2}], and [\ion{O}{1}] and their underlying continuum emission. We here only report on the continuum data and will report on the molecular and atomic line emission in a subsequent paper (Li et al. in preparation). The observations were carried out using Bands 4 and 6 with spatial resolutions of 0$\farcs$42 to 0$\farcs$65 with 36$-$44 12-m antennas. The on-source observation time was 5.5 hours and 3.4 hours, respectively. One of the 2 GHz spectral windows was tuned on the line and the other three on the continuum. The reference  redshift of $z =$ 6.0031 is from previous ALMA \cii\ observations presented in \citet{Wang2013}. The phase  and flux calibrators were  J2253+1608 and Pallas. The flux density scale calibration accuracy is better than 10$\%$. We reduced the data with the CASA pipeline. The noise levels of the dust maps are 0.02$-$0.04 mJy beam$^{-1}$.

\section{Results}
\label{red}

\begin{figure}
\centering
\subfigure{\label{J2310mom0}\includegraphics[scale=0.45]{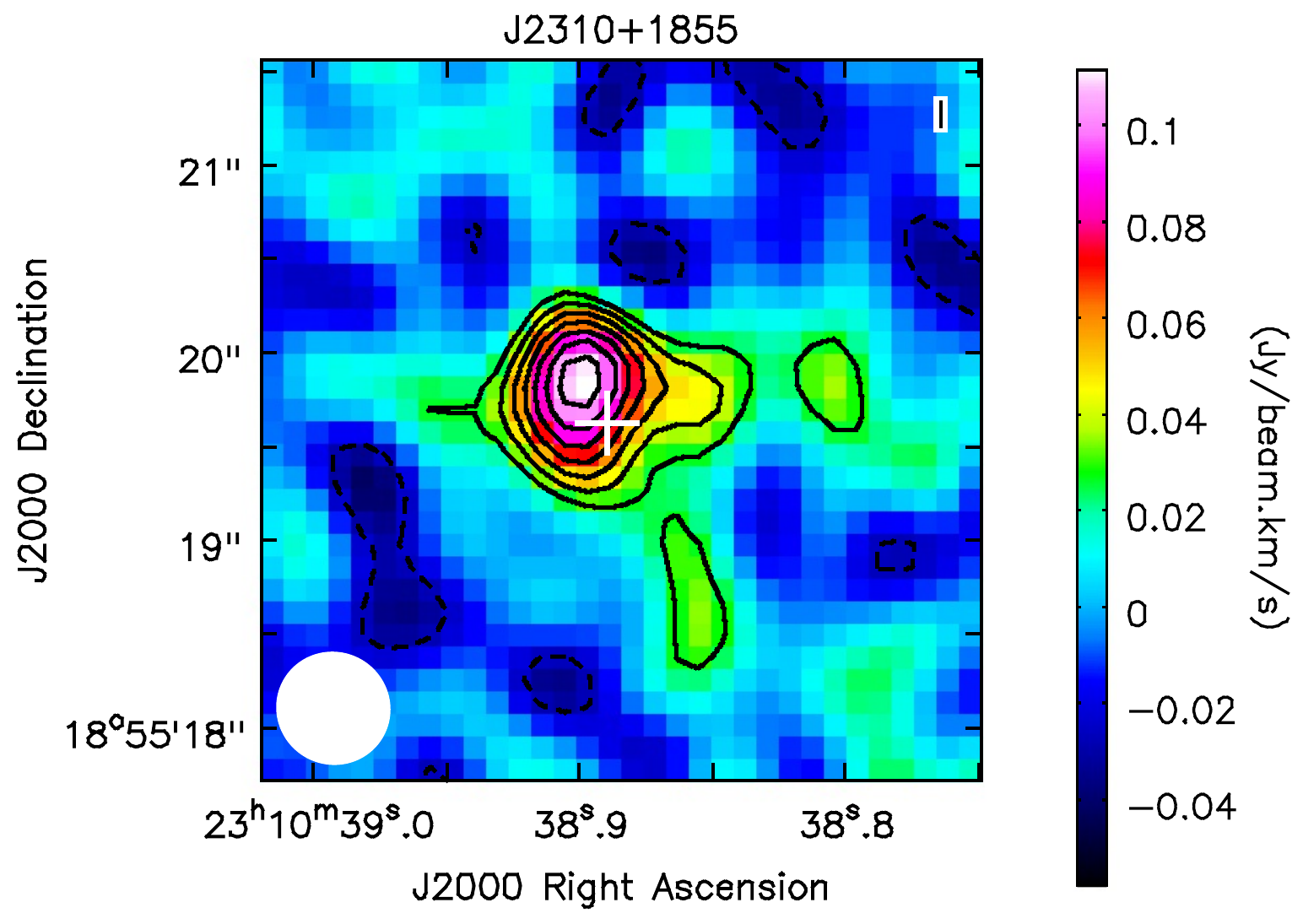}} 
\subfigure{\label{vm}\includegraphics[scale=0.42]{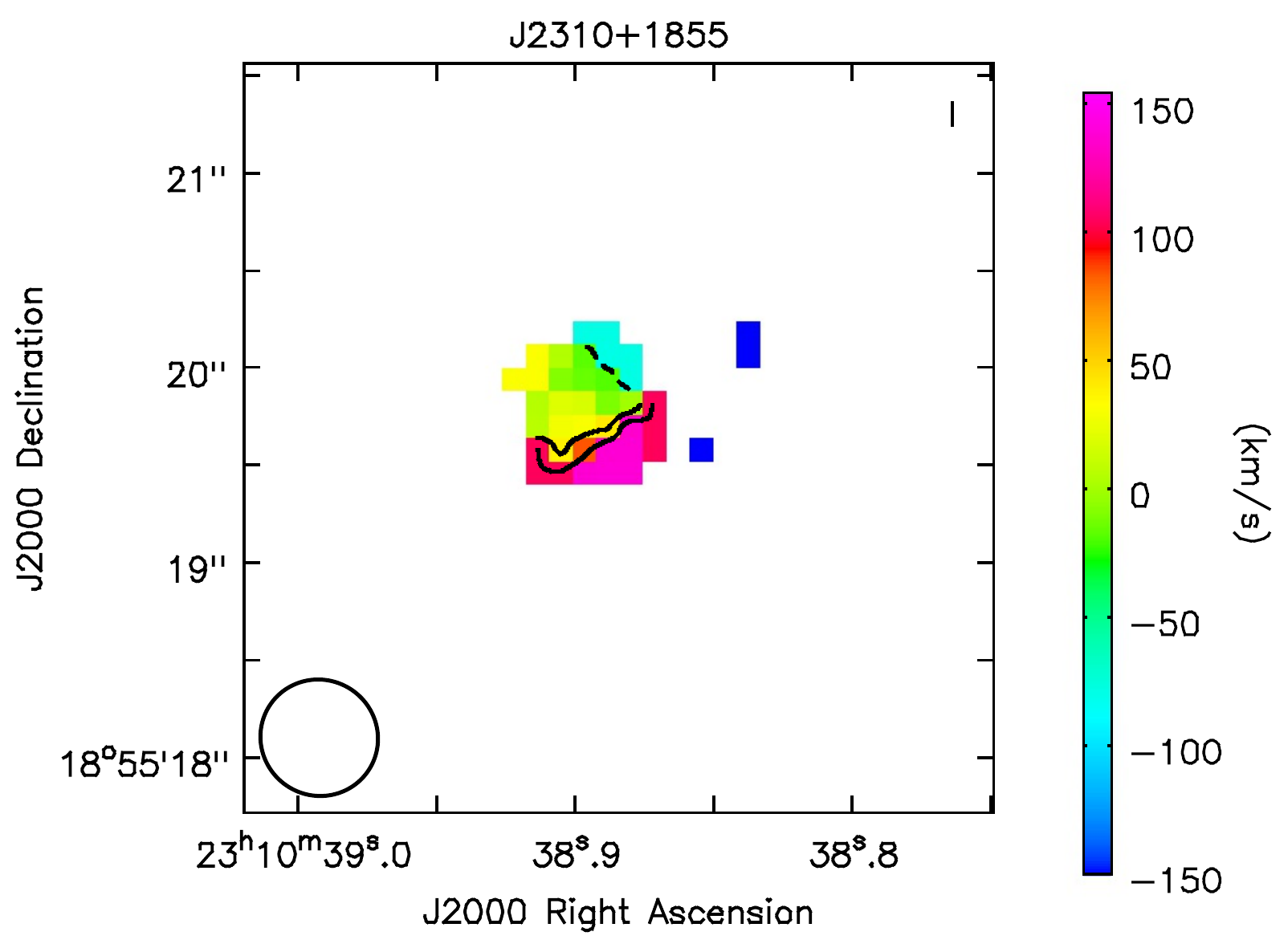}}
\subfigure{\label{J0129mom0}\includegraphics[scale=0.43]{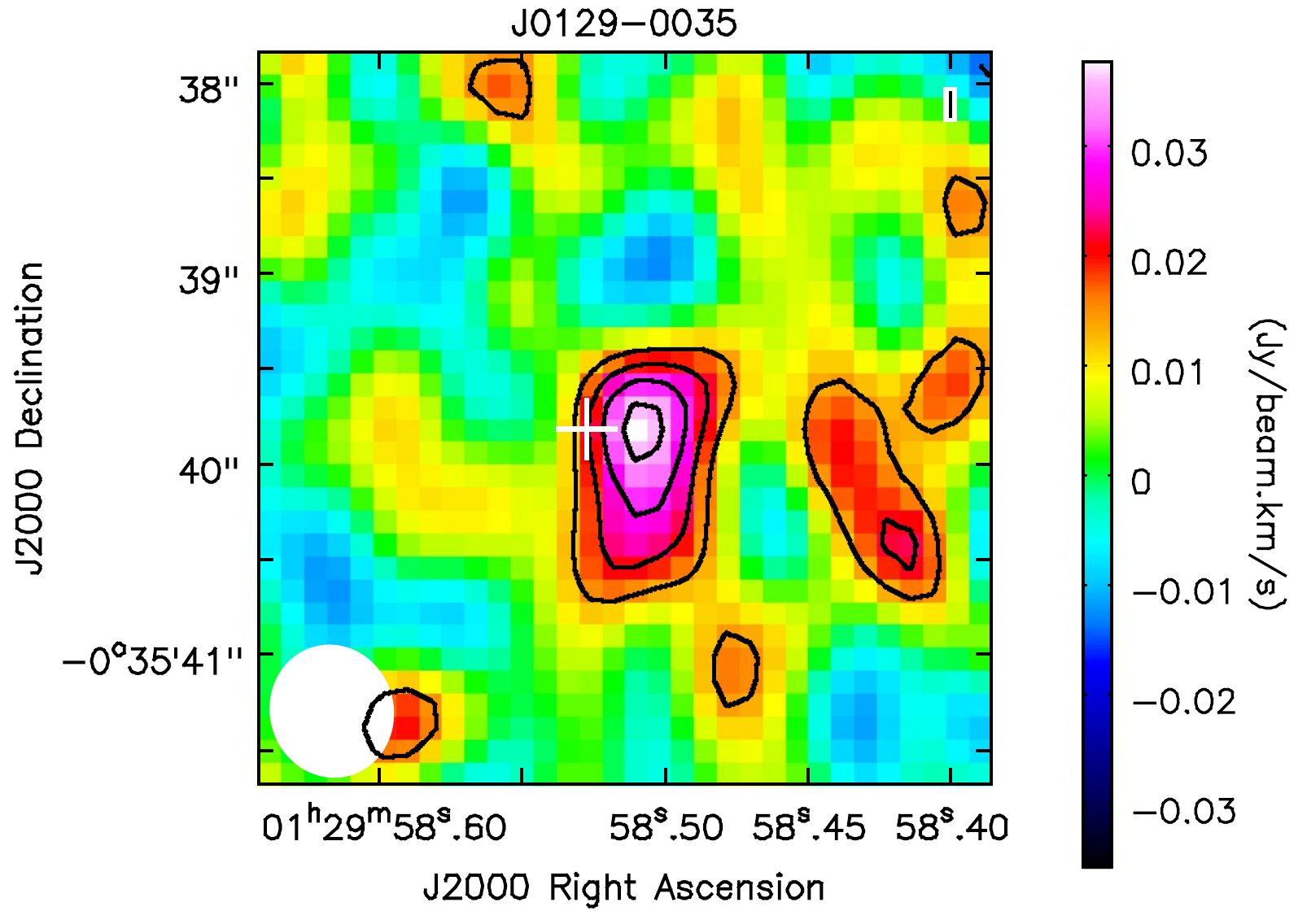}} 
\subfigure{\label{J2054mom0}\includegraphics[scale=0.43]{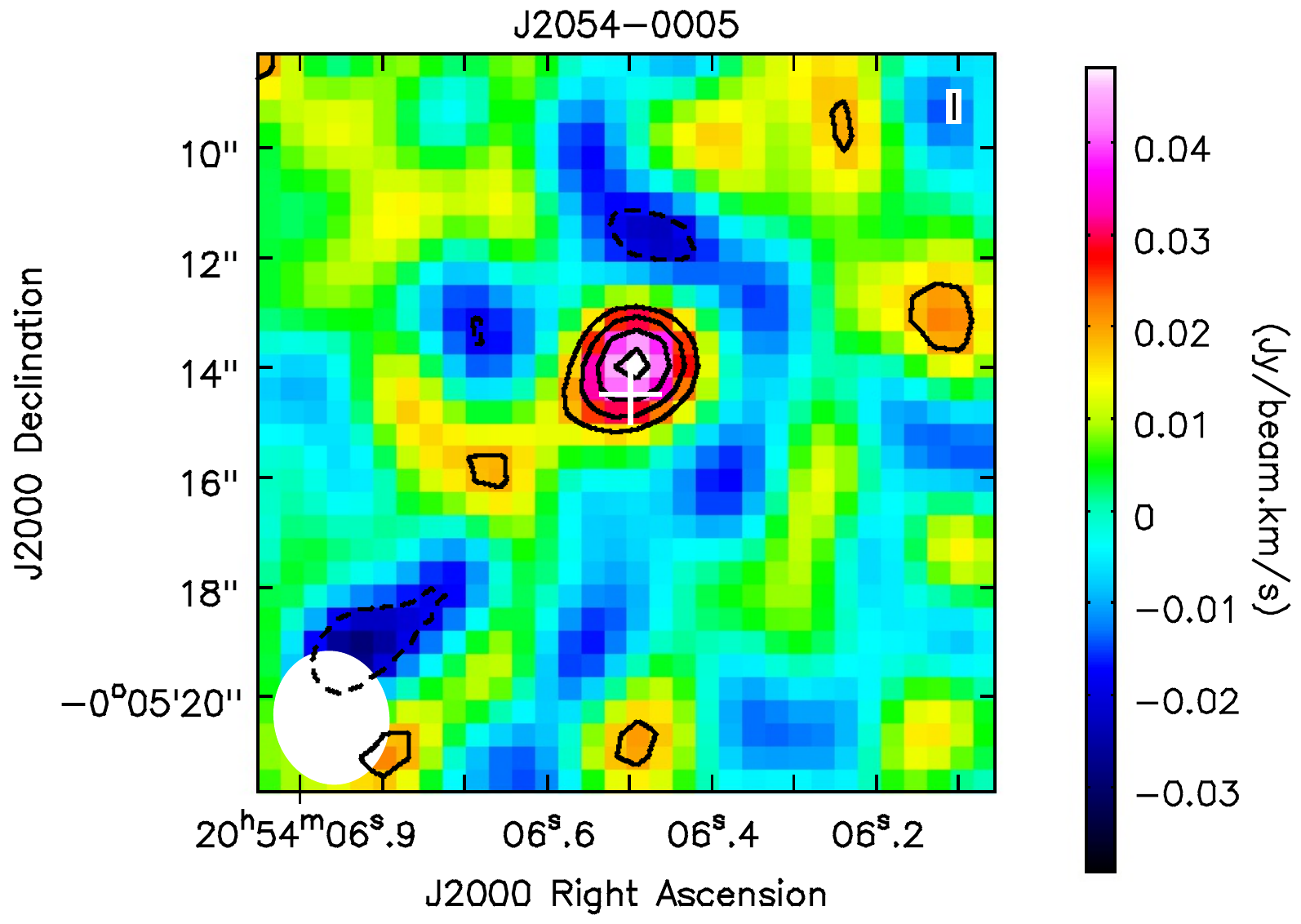}} 
\caption{Top row from left to right:  the CO (2$-$1) velocity-integrated map and velocity map produced with an intensity cut above 3.5$\sigma$ of J2310+1855. Bottom row from left to right: the CO (2$-$1) velocity-integrated maps of J0129$-$0035 and J2054$-$0005. The white crosses show the quasar optical positions  calibrated by Gaia astrometry with nearby bright stars. The sizes of the synthesised beams are plotted in the bottom left of each panel: $0\farcs61$  $\times$  $0\farcs59$, $0\farcs69$  $\times$  $0\farcs64$, and $2\farcs42$  $\times$  $2\farcs08$ for J2310+1855, J0129$-$0035 and J2054$-$0005, respectively. Contour levels for each CO (2$-$1) intensity map are as follows: J2310$+$1855 - [$-$2, 2, 3, 4, 5, 6, 7, 8]  $\times$  13 mJy beam $^{-1}$ km s$^{-1}$, J0129$-$0035 - [$-$2, 2, 3, 4, 5]  $\times$  7 mJy beam $^{-1}$ km s$^{-1}$, J2054$-$0005 - [$-$2, 2, 3, 4, 5]  $\times$  9 mJy beam $^{-1}$ km s$^{-1}$. The contours in J2310+1855 velocity map are of [$-$1, 1, 2] $\times$ 50 km s$^{-1}$. Note that the unit contour levels of the intensity maps are the noise values for the three maps.  }
\label{co21mom0}
\end{figure}

\begin{figure}
\centering
\subfigure{\label{J2310all}\includegraphics[scale=0.31]{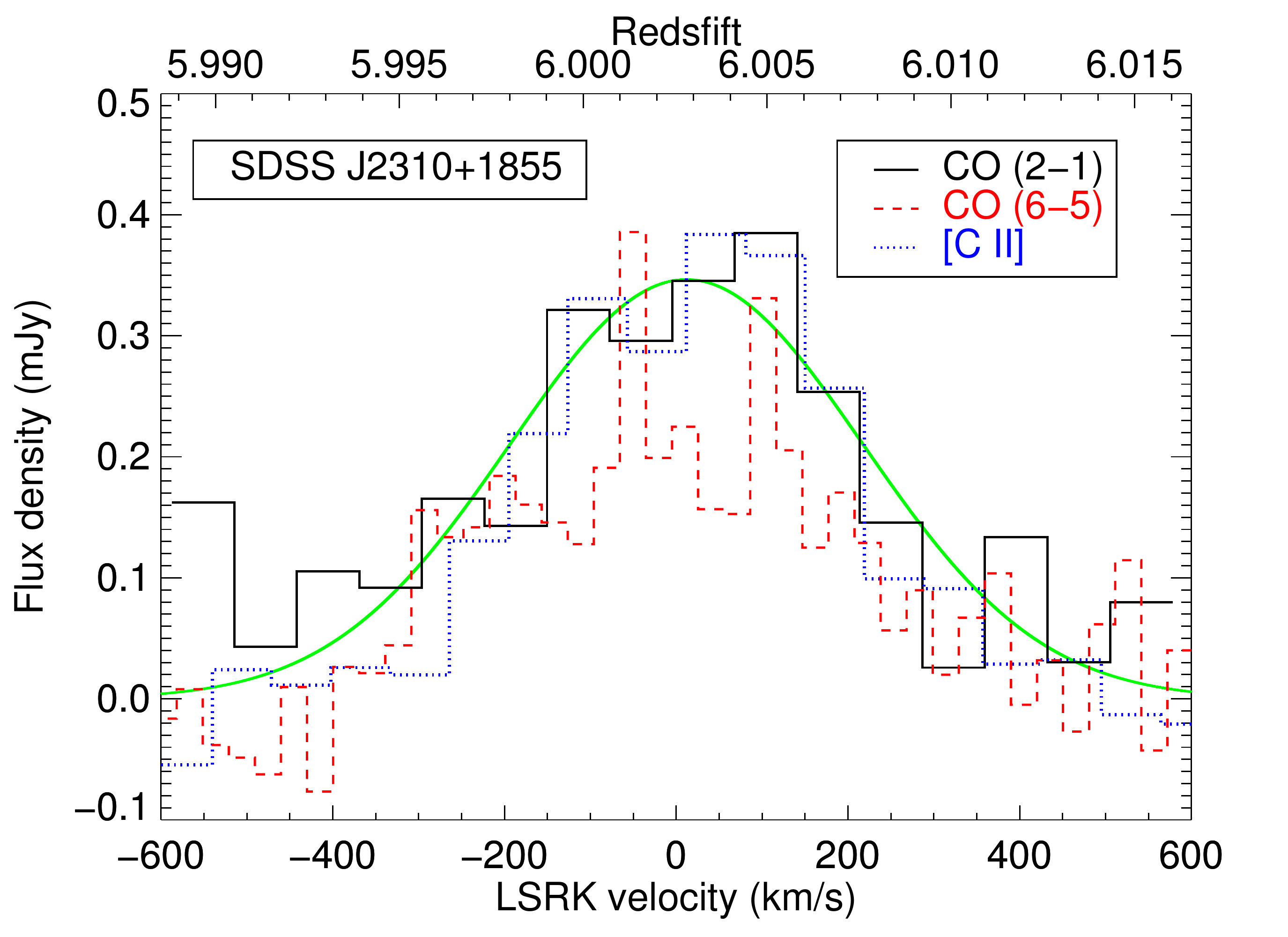}}
\subfigure{\label{J0129all}\includegraphics[scale=0.29]{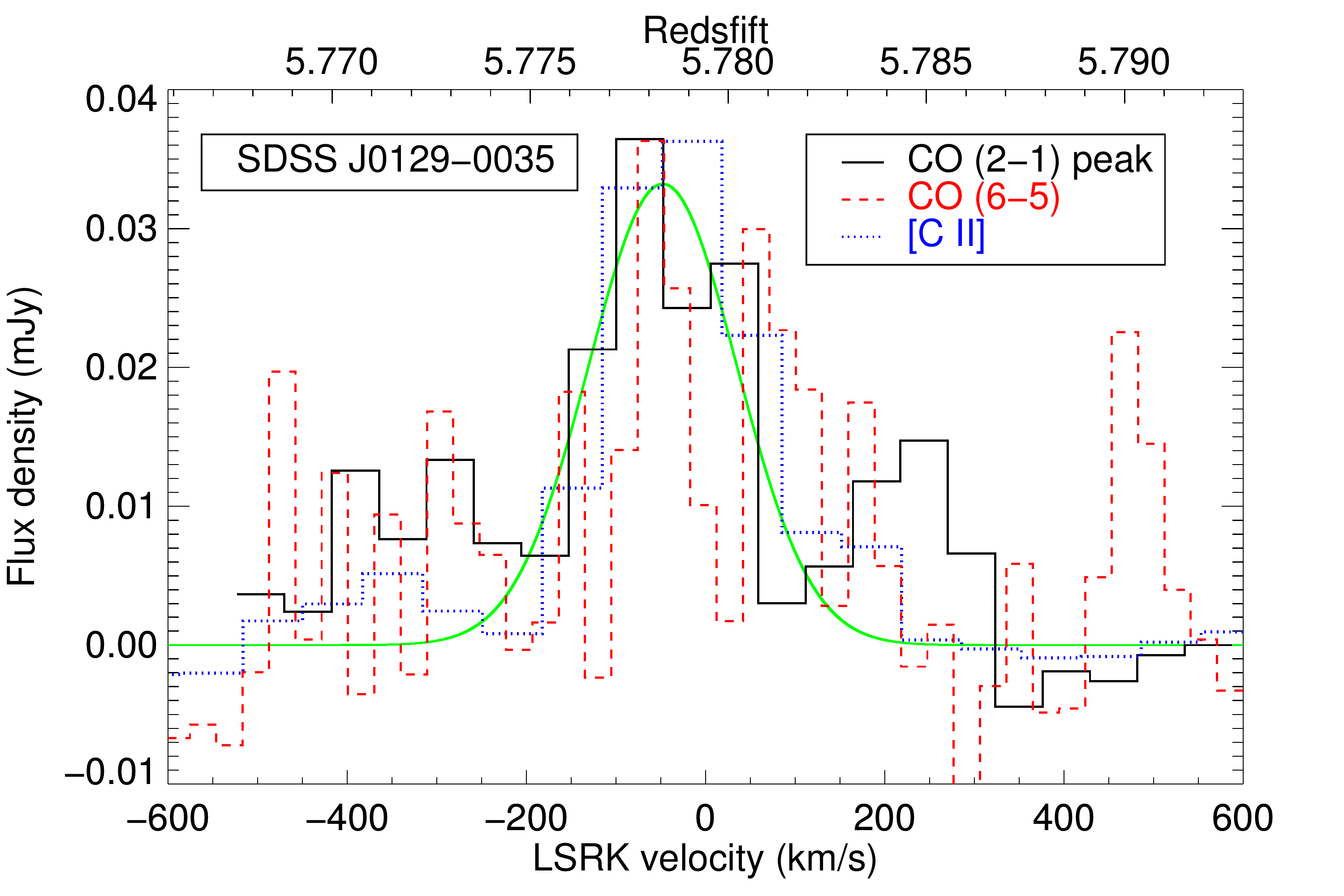}}
\subfigure{\label{J2054all}\includegraphics[scale=0.31]{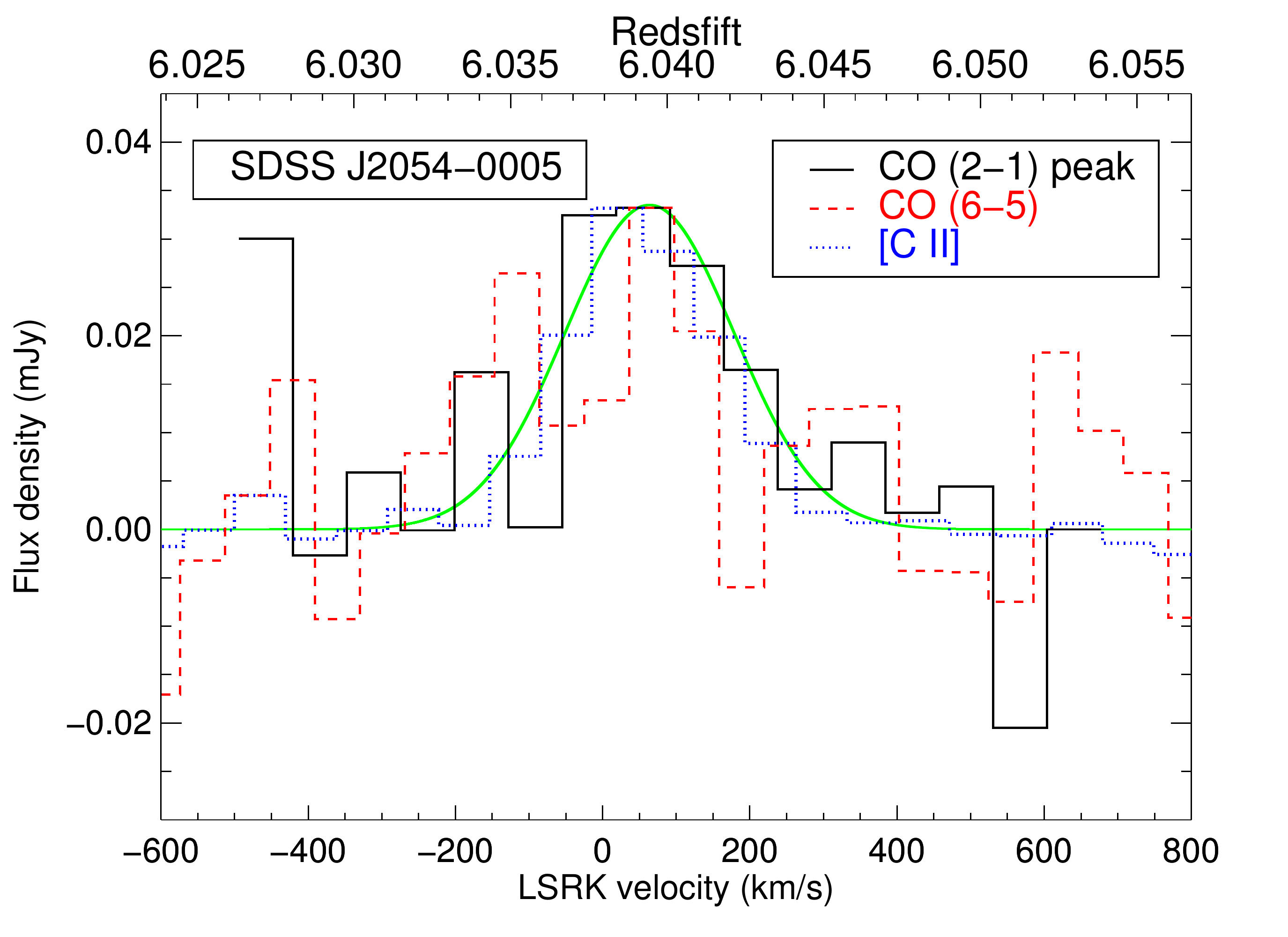}}
\caption{CO (2$-$1) line spectra of the three $z\sim6$ quasars, integrating the CO (2$-$1) data cube over all pixels detected at $>$ 2$\sigma$. The CO (2$-$1) line spectra (black solid line) are plotted over the emission lines of CO (6$-$5) - red-dashed line -  from \citet{Wang2010, Wang2011} and [\ion{C}{2}] - blue-dotted lines - from \citet{Wang2013}, which have been re-scaled to the peak of the CO (2$-$1). Gaussian fits to the CO (2$-$1) emission lines are shown as green solid lines. }
\label{cospectra}
\end{figure}

\subsection{J2310$+$1855}
This source was discovered in the SDSS (\citealt{Wang2013}; \citealt{Jiang2016}). With $m_{\rm 1450\AA}$ = 19.30 mag, it is one of the brightest optical quasars among the known $z\sim6$ quasars. The 250 GHz dust continuum flux density is  8.29 $\pm$ 0.63 mJy (\citealt{Wang2013}), which makes it one of the most FIR-luminous quasars known at $z\sim6$.  
The top left panel of Figure \ref{co21mom0} shows the CO (2$-$1) intensity map integrated from $-$333 to 323 km s$^{-1}$. The white cross is the quasar position calibrated with nearby bright stars by Gaia astrometry. The uncertainties of the Gaia calibrated position are 96 mas in RA and 106 mas in Dec. Using a 2D Gaussian to fit the CO (2$-$1) velocity integrated map, the source is found to be marginally resolved with a source size of ($0\farcs602$ $\pm$ $0\farcs184$) $\times$ ($0\farcs400$  $\pm$  $0\farcs208$). This size is consistent within the errors with that measured by [\ion{C}{2}] line emission (($0\farcs56$ $\pm$ $0\farcs03$) $\times$ ($0\farcs39$ $\pm$ $0\farcs04$); \citealt{Wang2013}). The corresponding physical source size is about (3.51 $\pm$ 1.07) kpc $\times$ (2.33  $\pm$  1.21) kpc, which is comparable to other $z\sim6$ quasars \citep{Wang2013}. The source position derived from the CO (2$-$1) line is consistent with these from [\ion{C}{2}] and its underlying dust continuum emission in \citet{Wang2013} and the Gaia calibrated position. We fitted a Gaussian profile to the CO (2$-$1) spectrum (left panel of Figure \ref{cospectra}), and measured a redshift of 6.0029  $\pm$  0.0005, which is in agreement with the redshifts from other ISM tracers (6.0025 $\pm$ 0.0007 from CO (6$-$5); 6.0031 $\pm$ 0.0002 from [\ion{C}{2}]; \citealt{Wang2013}). We fitted a line width (FWHM) of 484  $\pm$  48 km s$^{-1}$, which is consistent with the CO (6$-$5) line width (456 $\pm$ 64 km s$^{-1}$) but somewhat larger (2$\sigma$) than the [\ion{C}{2}] line width (393 $\pm$ 21 km s$^{-1}$). These measurements are given in Table \ref{comeas}. The velocity map of CO (2$-$1) line emission (top right panel of Figure \ref{co21mom0}) shows a velocity gradient, which is consistent with that from [\ion{C}{2}] \citep{Wang2013}.

J2310$+$1855 was detected in two Herschel PACS bands at $\sim5\sigma$ and in two SPIRE bands at $\sim3\sigma$. In the case of  the SPIRE 500 $\mu$m band observations, we give 3$\sigma$ value as an upper limit.  The photometric results are listed in Table \ref{conpho}.

\begin{figure}
\centering
\includegraphics[scale=0.6]{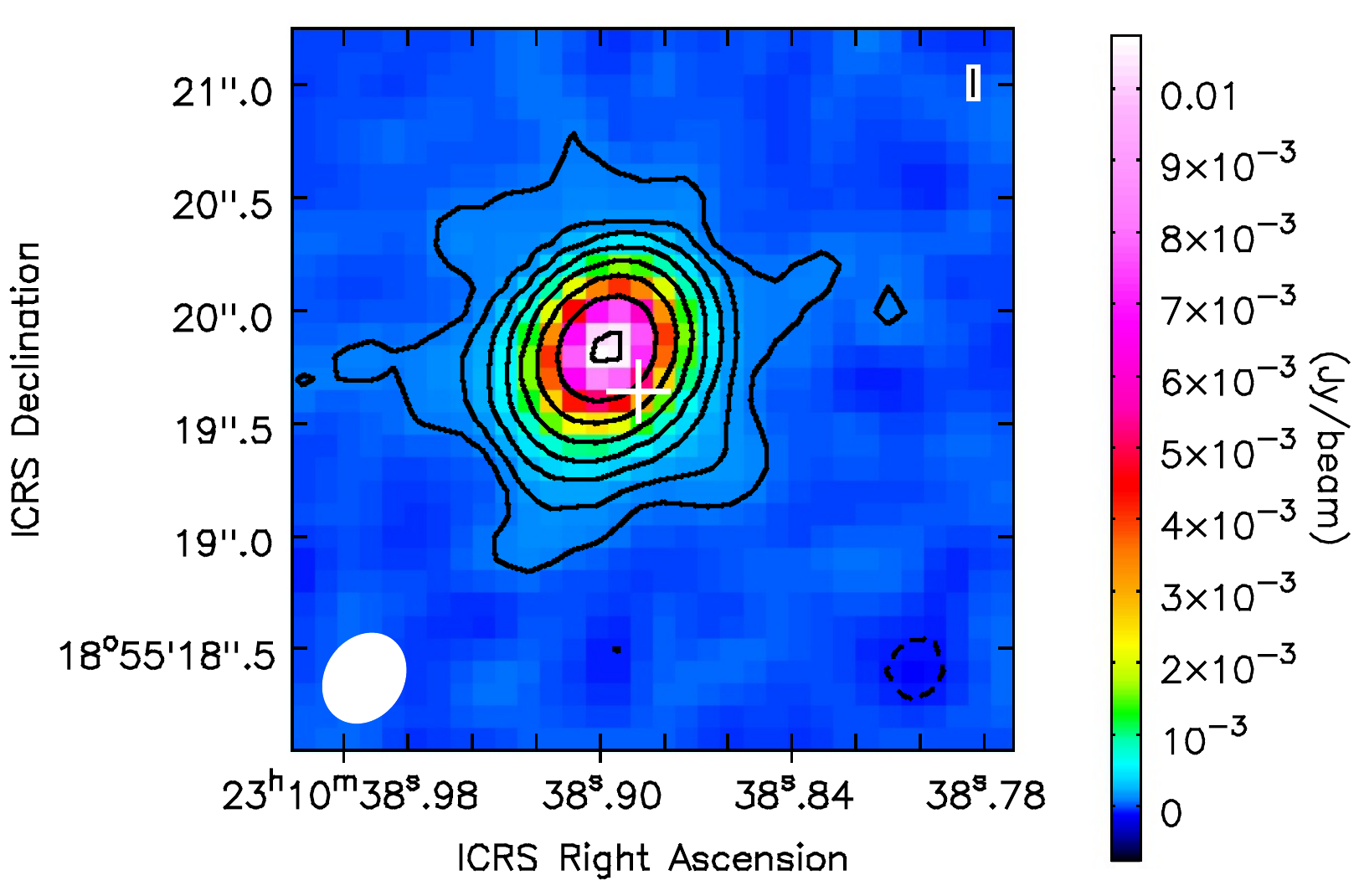}
\caption{The ALMA dust continuum map around 300 GHz of J2310+1855. The contours are of [$-$2, 2, 4, 8, 16, 32, 64, 128, 256] $\times$ 0.04 mJy beam$^{-1}$. The size of the synthesized beam (0$\farcs$42 $\times$ 0$\farcs$34) is shown in the lower left corner of the panel. The white cross represents the quasar position calibrated by the Gaia astrometry with nearby bright stars.}
\label{nii}
\end{figure}

The continuum flux densities using 2D Gaussian fits to the ALMA continuum maps at 140 GHz and 300 GHz are listed in Table \ref{conpho}. The continua are all spatially resolved at the observing frequencies, with consistent deconvolved source sizes of $\sim0\farcs2 \times 0\farcs2$. The derived source positions from the dust continua are consistent with that from CO (2$-$1) line. We present a dust continuum map at about 300 GHz as an example in Figure \ref{nii}.

\subsection{J0129$-$0035}
This quasar was selected from the SDSS stripe 82 with $m_{\rm 1450\AA}$ = 22.28 mag (\citealt{Jiang2009}), making it the faintest source among our targets. The flux density at 250 GHz is $2.73\pm0.49$ mJy \citep{Wang2011fir} yielding a FIR luminosity that is comparable to that of J2054$-$0005 but much smaller than that of J2310+1855.
The bottom left panel of Figure \ref{co21mom0} shows the CO (2$-$1) velocity-integrated map integrated from $-$126 to 85 km s$^{-1}$ with Gaia calibrated source position marked as white cross. The uncertainties of the Gaia calibrated position are 149 mas in RA and 114 mas in Dec (Wang et al. 2018 in preparation). A 2D Gaussian fit to the intensity map indicates that it is a point source. The inferred source position from the CO (2$-$1) line is marginally consistent with those from [\ion{C}{2}] and its underlying dust continuum emission \citep{Wang2013} and the Gaia calibrated position. A Gaussian fit to the CO (2$-$1) peak spectrum provides a redshift of 5.7783  $\pm$  0.0004 and a  line width of FWHM = 204 $\pm$ 45 km s$^{-1}$, and these results are consistent with measurements derived from CO (6$-$5) and [\ion{C}{2}] (\citealt{Wang2011fir}; \citealt{Wang2013}). 

We did not detect J0129$-$0035 in any of the Herschel SPIRE bands. The 3$\sigma$ values upper limits are listed in Table \ref{conpho}.

\subsection{J2054$-$0005}
This source was discovered by \citet{Jiang2008} from SDSS stripe 82 with $m_{\rm1450\AA}$ = 20.60 mag. 
The bottom right panel of Figure \ref{co21mom0} presents the CO (2$-$1) intensity map, integrated over the velocity range from $-$18 to 201 km s$^{-1}$, where we also plotted the Gaia calibrated position as white cross. The uncertainties of the Gaia calibrated position are 25 mas in RA and 27 mas in Dec. We fitted a 2D Gaussian to the CO (2$-$1) velocity-integrated map and found that it is a point source with a peak value of  0.055 $\pm$ 0.008 Jy km s$^{-1}$ beam$^{-1}$. The derived CO (2$-$1) source position is 20h54m06.5001s $\pm$ 284 mas of RA and $-$00d05m14.0622s $\pm$ 267 mas of Dec, where the errors are the sum of the fitting-type error and the position error caused by thermal noise \citep{Reid2014}. This position is $\sim0\farcs4$ away from those inferred from [\ion{C}{2}] (position errors are $\sim$ 20 mas in both RA and Dec) and its underlying dust continuum emission  (position errors are $\sim$ 10 mas in both RA and Dec) in \citet{Wang2013} and the Gaia calibrated position. High sensitivity observations of the molecular CO lines are needed to check this tentative offset. A Gaussian fit to the line spectrum yielded a redshift of 6.0394 $\pm$ 0.0004, consistent with CO (6$-$5) and [\ion{C}{2}] measurements (6.0379  $\pm$  0.0022 and 6.0391  $\pm$  0.0001 respectively; \citealt{Wang2010}; \citealt{Wang2013}).

\begin{deluxetable}{lcccccc}
\tabletypesize{\scriptsize}  
\tablewidth{0pt} 
\tablecaption{CO (2$-$1) results compared with previous work\label{comeas}}
\tablecolumns{7}
\tablehead{
\colhead{Species } & \colhead{Redshift} & \colhead{FWHM} & \colhead{Flux}& \colhead{Luminosity } &\colhead{$L^{'}_{\rm CO (1-0)}$}& \colhead{$M_{\rm gas}$} \\
\colhead{ } & \colhead{ } & \colhead{(km s$^{-1}$)} & \colhead{(Jy km s$^{-1}$)}& \colhead{(10$^{6}$ $L_{\odot}$)} & \colhead{(10$^{10}$$\rm K \ km \ s^{-1} pc^{2}$)}& \colhead{(10$^{10}$$M_{\odot}$)} \\
\colhead{(1)} & \colhead{(2)} & \colhead{(3)} & \colhead{(4)}& \colhead{(5)}& \colhead{(6)}  & \colhead{(7)} 
}
\startdata
J2310+1855 &  &  &  &  & & \\
CO (6$-$5) &6.0025 $\pm$ 0.0007 &456 $\pm$ 64 &1.52 $\pm$ 0.13  &543.0 $\pm$ 46.4&6.5 $\pm$ 0.6&5.2\\
\cii\  &6.0031 $\pm$ 0.0002 &393 $\pm$ 21 &8.83 $\pm$ 0.44 &8700 $\pm$ 1400 &-&- \\
CO (2$-$1) &6.0029 $\pm$ 0.0005 &484 $\pm$ 48 &0.18 $\pm$ 0.02 &21.3 $\pm$ 0.2 &5.4 $\pm$ 0.5&4.3   \\
\tableline
J0129$-$0035 &  & &  & &&\\
CO (6$-$5) &5.7794 $\pm$ 0.0008 &283 $\pm$ 87 &0.37 $\pm$ 0.07  &125.0 $\pm$  23.6&1.5 $\pm$ 0.3&1.2\\
\cii\ &5.7787 $\pm$ 0.0001 &194 $\pm$ 12 &1.99 $\pm$ 0.12  &1800 $\pm$ 300 &-&- \\
CO (2$-$1) peak &5.7783  $\pm$  0.0004 &195 $\pm$ 41 &0.036 $\pm$ 0.005 & 4.0 $\pm$ 0.6 &1.0 $\pm$ 0.1&0.8\\
\tableline
J2054$-$0005 & & & & &&\\
CO (6$-$5) &6.0379 $\pm$ 0.0022 &360 $\pm$ 110 &0.34 $\pm$ 0.07 &122.5 $\pm$ 25.2  &1.5 $\pm$ 0.3&1.2\\
\cii\ &6.0391 $\pm$ 0.0001 &243 $\pm$ 10  &3.37 $\pm$ 0.12  &3300 $\pm$ 500 &-&- \\
CO (2$-$1) peak &6.0394 $\pm$ 0.0004 &270 $\pm$ 47 &0.06 $\pm$ 0.01 &6.6 $\pm$ 0.9  &1.7 $\pm$ 0.2&1.3
\enddata
\tablerefs{CO (6$-$5) information comes from \citet{Wang2010} and \citet{Wang2011fir}, and \cii\ information is from \citet{Wang2013}. }
\tablecomments{Column 1: different transitions. Column 2: redshifts from different ISM tracers. Column 3: FWHM of fitted Gaussian profile. Column 4: line flux by integrating the fitted Gaussian profile or the peak flux  in units of Jy km s$^{-1}$ beam$^{-1}$ from a 2D Gaussian fit to the intensity map. Column 5$-$6: line luminosities following the method in \citet{Solomon1992}. Column 7: molecular gas mass with a conversion factor $\alpha_{\rm CO}\sim0.8 \ M_{\odot} \ (\rm K\ km\ { s }^{ -1 }\ { pc }^{ 2 })^{-1} $ and assuming $L^{' }_{\rm CO (2-1)} \approx L^{' }_{\rm CO (1-0)}$.}
\end{deluxetable}

\begin{deluxetable}{lcccc}
\tabletypesize{\scriptsize}
\tablecaption{Continuum photometry information\label{conpho}}
\tablewidth{0pt}
\tablecolumns{5}
\tablehead{
\colhead{ } &\colhead{Telescope } & \colhead{J2310$+$1855} & \colhead{J0129$-$0035}  & \colhead{J2054$-$0005} 
 }
\startdata
$m_{\rm 1450\AA}$ (mag)       &-&19.30$^{g}$&22.28$^{e}$&20.6$^{c}$\\
$z$ (mag)                   &SDSS&19.31 $\pm$ 0.11$^{a}$&22.16 $\pm$ 0.11$^{e}$&20.72 $\pm$ 0.09$^{c}$\\
$w_{1}$ (mag)           &WISE&18.48 $\pm$ 0.05$^{b}$&-&-\\
$w_{2}$ (mag)           &WISE&18.73 $\pm$ 0.12$^{b}$&-&-\\
$w_{3}$ (mag)           &WISE&17.51 $\pm$ 0.44$^{b, 1}$&-&-\\
$F_{\rm 100 \mu m}$ (mJy)     &Herschel/PACS&\textbf{6.5 $\pm$ 1.2}&-&$<$ 2.7$^{h}$\\
$F_{\rm 160 \mu m}$ (mJy)    &Herschel/PACS&\textbf{13.2 $\pm$ 2.8} &-&9.8 $\pm$ 1.3$^{h}$\\
$F_{\rm 250 \mu m}$ (mJy)    &Herschel/SPIRE& \textbf{19.9 $\pm$ 6.0}&\textbf{$<$ 12.2}&15.2 $\pm$ 5.4$^{h}$\\
$F_{\rm 350 \mu m}$ (mJy)    &Herschel/SPIRE&\textbf{22.0 $\pm$ 6.9}&\textbf{$<$ 11.4}&12.0 $\pm$ 4.9$^{h}$\\
$F_{\rm 500 \mu m}$ (mJy)    &Herschel/SPIRE&\textbf{$<$ 29.4}&\textbf{$<$ 15.2}&$<$ 19.5$^{h}$\\
$F_{496\rm GHz}$ (mJy)    &ALMA& 24.89 $\pm$ 0.72$^{j}$&-&-\\
$F_{350\rm GHz}$ (mJy)    &ALMA&\textbf{14.54 $\pm$ 0.21}&-&-\\ 
$F_{338\rm GHz}$ (mJy)    &ALMA&\textbf{14.49 $\pm$ 0.21}&-&-\\ 
$F_{295\rm GHz}$ (mJy)    &ALMA&\textbf{11.56 $\pm$ 0.10}&-&-\\ 
$F_{283\rm GHz}$ (mJy)    &ALMA&\textbf{11.60 $\pm$ 0.10}&-&-\\ 
$F_{263\rm GHz}$ (mJy)    &ALMA&8.91 $\pm$ 0.08$^{g}$& 2.57 $\pm$ 0.06$^{g}$& 2.98 $\pm$ 0.05$^{g}$\\
$F_{\rm 250 GHz}$ (mJy)    &IRAM&8.29 $\pm$ 0.63$^{g}$&2.37 $\pm$ 0.49$^{f}$&2.38 $\pm$ 0.53$^{d}$\\
$F_{147\rm GHz}$ (mJy)    &ALMA&\textbf{1.59 $\pm$ 0.04}&-&-\\ 
$F_{143\rm GHz}$ (mJy)    &ALMA&\textbf{1.41 $\pm$ 0.04}&-&-\\ 
$F_{135\rm GHz}$ (mJy)    &ALMA&\textbf{1.24 $\pm$ 0.04}&-&-\\ 
$F_{131\rm GHz}$ (mJy)    &ALMA&\textbf{1.10 $\pm$ 0.05}&-&-\\ 
$F_{91.5\rm GHz}$ (mJy)    &ALMA& 0.41 $\pm$ 0.03$^{i}$&-&-\\
\enddata
\tablerefs{$^{a}$SDSS; $^{b}$WISE ; $^{c}$\citet{Jiang2008}; $^{d}$\citet{Wang2008thermal};  $^{e}$\citet{Jiang2009}; $^{f}$\citet{Wang2011}; $^{g}$\citet{Wang2013}; $^{h}$\citet{Leipski2014}; $^{i}$\citet{Feruglio2018}; $^{j}$\citet{Hashimoto2018}.}
\tablecomments{Column 1 indicates the different continuum bands. Column 2 is the telescope that did the corresponding observations. Columns 3$-$5 represent the three sources. The detections in boldface are from our work, and the quoted upper limits are 3$\sigma$. All magnitudes are in AB magnitude after correcting for Milky Way extinction.}
\end{deluxetable}

\begin{figure}
\centering
\includegraphics[scale=0.5]{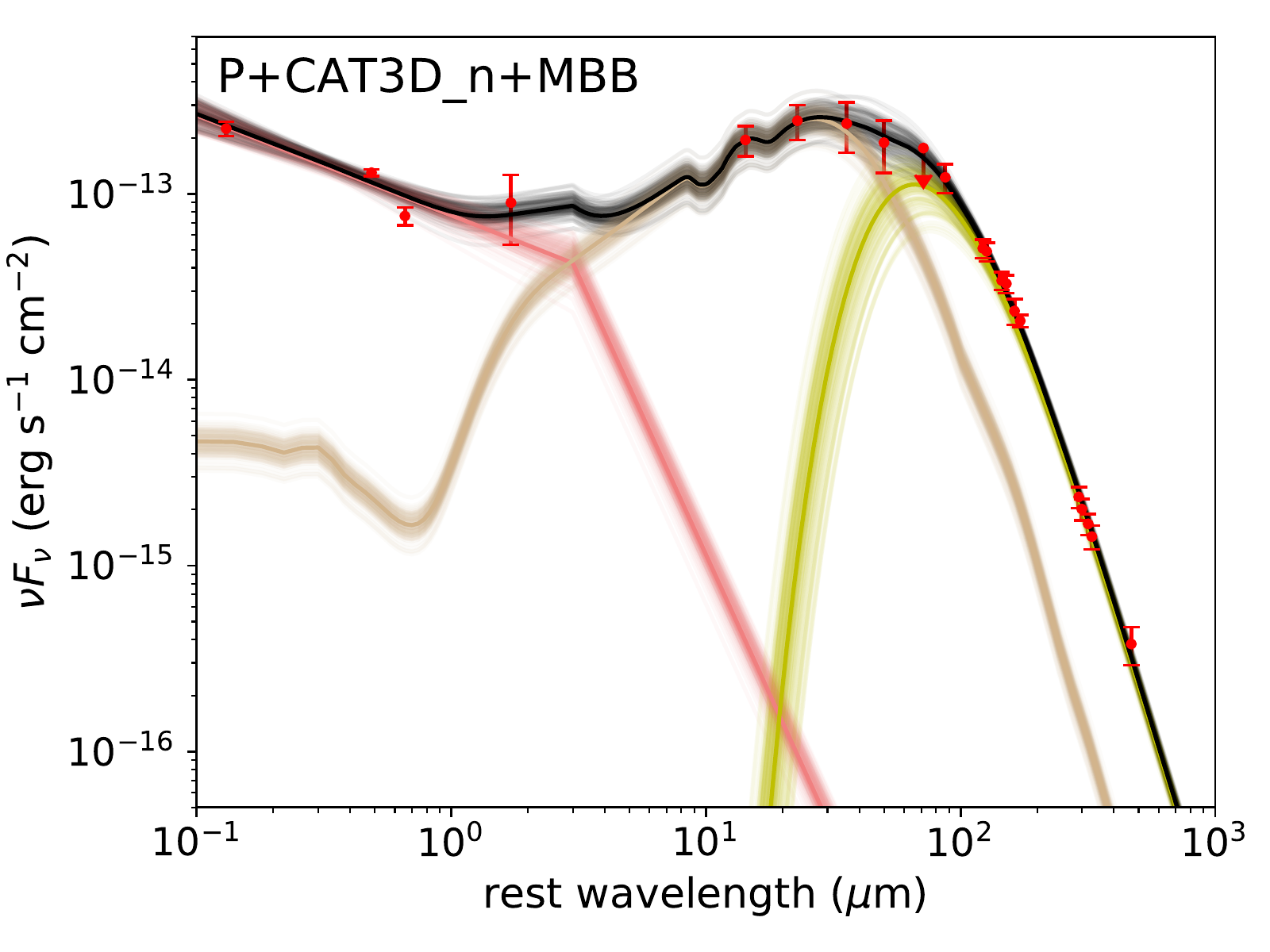}
\includegraphics[scale=0.5]{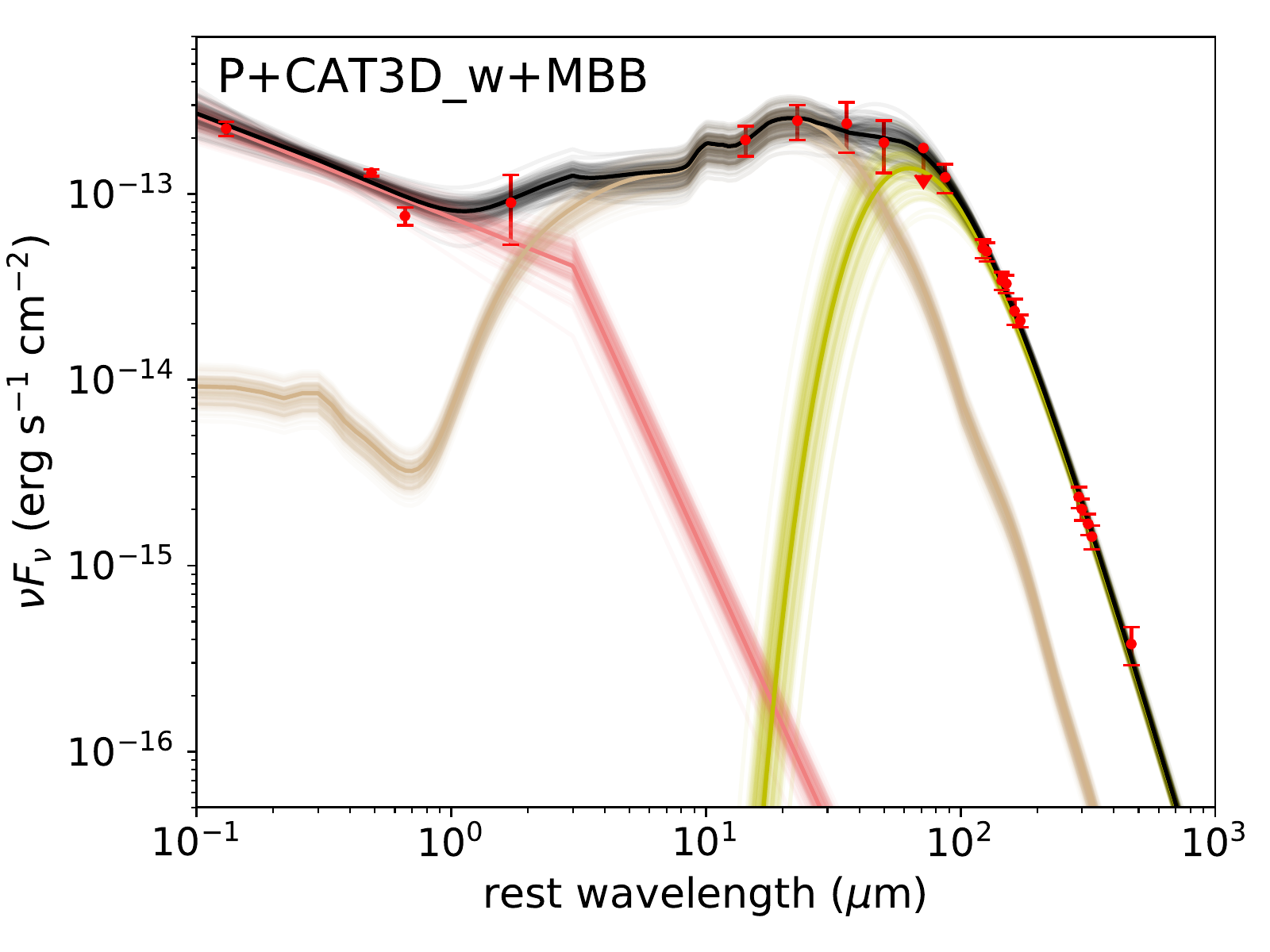}
\includegraphics[scale=0.5]{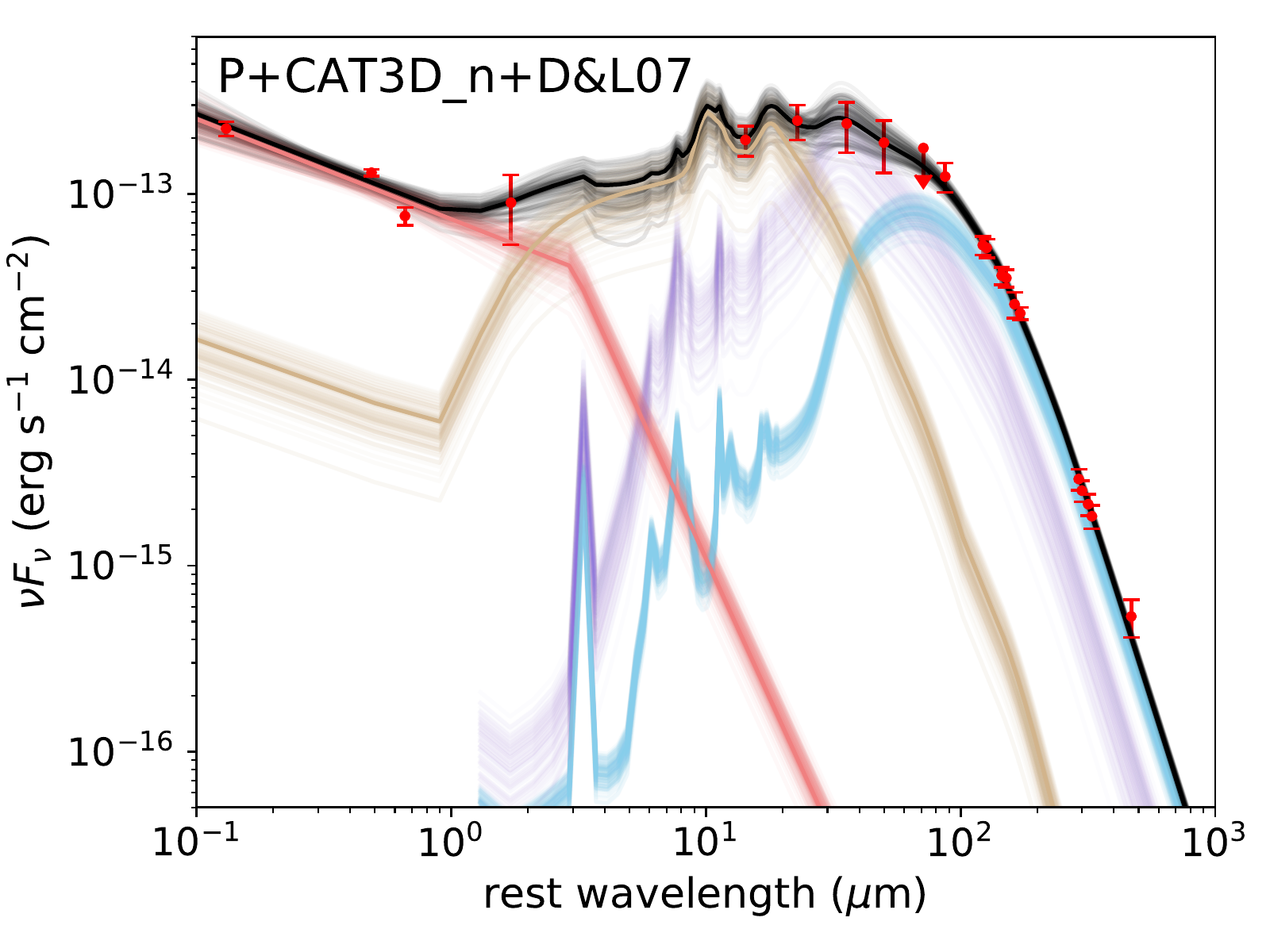}
\includegraphics[scale=0.5]{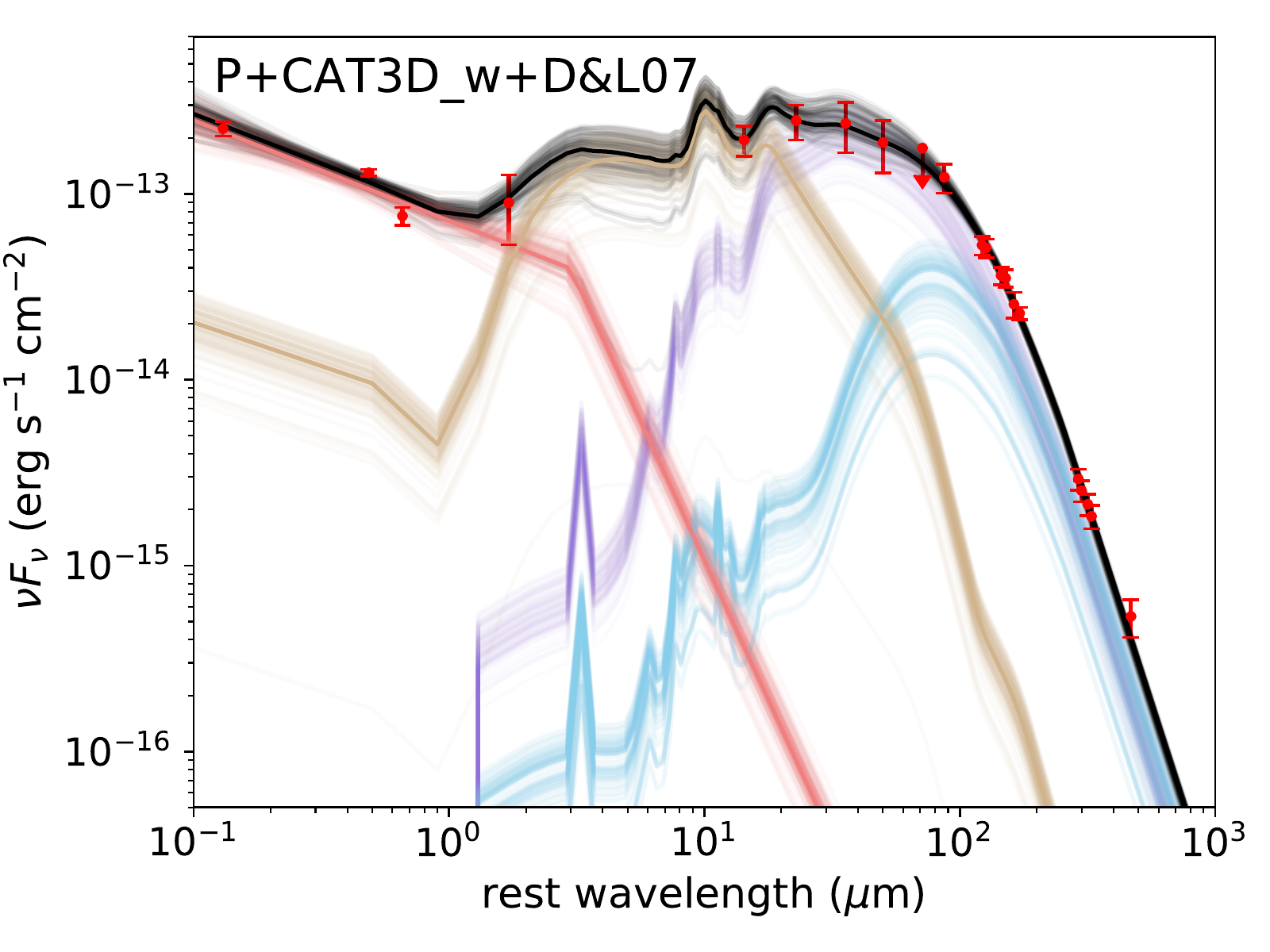}
\caption{SED fit for SDSS J2310+1855 with different components. The four panels indicate four models as labeled in the top right of each panel. `P' represents a UV/optical power law. `CAT3D$\_$n' and `CAT3D$\_$w' represent CAT 3D  AGN dust torus models without and with wind. `MBB' is a FIR modified black body. `D$\&$L07' refers to the dust model in \citet{Draine2007}. The red points with error bars or downward arrows are observed data. The pink lines represent a UV/optical power law from an accretion disk. The brown lines are from the CAT3D AGN torus model. The green lines correspond to a MBB profile heated by the star formation activity. The purple and blue lines present emissions from PDR and diffuse regions defined in \citet{Draine2007}. The black lines are the sum of all components.}
\label{sedfits}
\end{figure}

\begin{deluxetable}{lccccc}
\tabletypesize{\scriptsize}  
\tablewidth{0pt} 
\tablecaption{Physical properties derived from the SED fit of J2310$+$1855\label{sedfitsmea}}
\tablecolumns{6}
\tablehead{
\colhead{ } &\colhead{ }& \colhead{P+CAT 3D$\_$n+MBB} & \colhead{P+CAT 3D$\_$w+MBB}  & \colhead{P+CAT 3D$\_$n+D$\&$L07} & \colhead{P+CAT 3D$\_$w+D$\&$L07}
}
\startdata
$\alpha_{\rm{UV/opt}}$ &(1)&$-$0.47$_{-0.08}^{+0.08}$  &$-$0.46$_{-0.08}^{+0.09}$  &$-$0.47$_{-0.08}^{+0.08}$ &$-$0.47$_{-0.09}^{+0.08}$ \\
$L_{\rm{UV/opt}}$ (10$^{46}$ erg s$^{-1}$) &(2)& 14.6$_{-1.0}^{+1.0}$ & 14.4$_{-1.0}^{+1.0}$  & 14.0$_{-1.0}^{+1.0}$ & 13.8$_{-1.0}^{+1.0}$ \\
$a$                               &(3)&$-$1.25&$-$1.50&$-$1.75 &$-$3.00\\
$h$                               &(4)&1.25&0.50&0.25 &0.10\\
$N_{0}$                        &(5)&10.0&10.0&5.0 &10.0\\
$a_{\rm{w}}$                &(6)&-&$-0.50$&- &$-$0.50\\
$\theta_{\rm{w}}$         &(7)&-&30&- &30\\
$\sigma_{\rm{\theta}}$ &(8)&-&7.50&- &10.00\\
$f_{\rm{wd}}$               &(9)&-&2.00&-&0.15\\
$R_{\rm{out}}$             &(10)&500.0&500.0&500.0 &450.0\\
$\tau_{\rm{cl}}$            &(11)&50.0&50.0&50.0 &50.0\\
Inclination (degree)      &(12)&15.0&0.0&0.0 &45.0\\
$\gamma$&(13)& - & -   & 0.17$_{-0.07}^{+0.07}$ & 0.39$_{-0.11}^{+0.13}$ \\
$U_{\rm min}$&(14)& - & -   &25.0  &25.0  \\
$U_{\rm max}$&(15)& - & -   &10$^{6}$  &10$^{6}$  \\
$q_{\rm PAH}$ (\%)&(16)& - & -   &0.47&0.01  \\
$T_{\rm{dust}}$ (K)    &(17)& 39$_{-3}^{+3}$ & 40$_{-2}^{+3}$   &$\bf36_{-2}^{+2}$&$\bf40_{-2}^{+2}$\\
$L_{\rm{FIR}}$ ($10^{13}L_{\odot}$) &(18)& $\bf1.4_{-0.3}^{+0.3}$   & $\bf1.6_{-0.3}^{+0.3}$     & $\bf3.2_{-0.7}^{+0.7}$ & $\bf3.6_{-0.6}^{+0.6}$ \\
SFR ($10^{3}M_{\odot} \rm yr^{-1}$) &(19)& $\bf2.4_{-0.4}^{+0.5}$   & $\bf2.7_{-0.5}^{+0.6}$     & $\bf5.5_{-1.2}^{+1.1}$  & $\bf6.1_{-1.0}^{+1.0}$ \\
$M_{\rm{dust}}$ ($10^{9}M_{\odot}$)&(20)& $\bf1.7_{-0.3}^{+0.4}$ & $\bf1.6_{-0.3}^{+0.3}$   &3.8$_{-0.3}^{+0.2}$&3.9$_{-0.3}^{+0.3}$ \\
\enddata
\tablecomments{Column 1: SED fit parameters. Column 2$-$5: different SED models for SDSS J2310+1855, where `P' is UV/opt power law, `CAT 3D$\_$n' presents CAT 3D dust torus model without a polar wind, `CAT 3D$\_$w' shows CAT 3D dust torus model with a polar wind, `MBB' represents FIR modified black body and `D$\&$L07' is \citet{Draine2007} dust model. Row 1: UV/opt power law slope defined as $F_{\nu} \varpropto \nu^{\alpha}$. Row 2: UV/opt luminosity determined by integrating the power law component between 0.1 $\mu$m  and 1 $\mu$m. Row 3$-$5: CAT 3D torus/disk parameters. $a$, $h$ and $N_{0}$ are the index of the radial dust cloud distribution power law, the torus dimensionless scale height and the number of clouds along an equatorial line-of-sight, respectively. Row 6$-$8: CAT 3D wind parameters (only for `CAT3D$\_$w'). $a_{\rm{w}}$, $\theta_{\rm{w}}$ and $\sigma_{\rm{\theta}}$ are the index of the dust cloud distribution power law along the wind, the half-opening angle of the wind and the angular width of the hollow wind cone, respectively. Row 9$-$11: CAT 3D global parameters. $f_{\rm{wd}}$, $R_{\rm{out}}$ and $\tau_{\rm{cl}}$ are the ratio of numbers of dust clouds along the line-of-sight of the wind to the dust clouds in the disk plane, the outer radius of the torus and the optical depth of the individual clouds, respectively. Row 12: the torus inclination. Row 13$-$16: parameters from the \citet{Draine2007} dust model. $\gamma$, $U_{\rm min}$, $U_{\rm max}$ and $q_{\rm PAH}$ are the PDR fraction, the minimum and maximum starlight intensity relative to the local interstellar radiation field, and the dust mass fraction in PAHs. Row 17: dust temperature. The first two columns are MBB temperature from the fitting procedure and the last two columns are the dust temperature with dust grain size $\geq$0.03 $\mu$m calculated by Equation \ref{dldt2}. Row 18: FIR luminosity determined by integrating the FIR dust model (the MBB or the \citet{Draine2007} dust model) between  8 $\mu$m  and 1000 $\mu$m. Row 19: SFR derived from the formula in \citet{Kennicutt1998}. Row 20: dust mass.  The first two columns are derived from Equation \ref{eq1}, and the last two columns are from the model fit. The values in boldface are derived from the fitted parameters which are in normal-type.}
\end{deluxetable}

\section{Analysis and Discussion}
\label{dis}

\subsection{Gas distribution and gas mass in quasar host galaxies}
\label{codis}

J0129$-$0035 and J2054$-$0005 are point sources in the CO (2$-$1) line emission, and J2310$+$1855 is marginally resolved with a physical size of (3.51 $\pm$ 1.07) kpc $\times$ (2.33  $\pm$  1.21) kpc. Figure \ref{cospectra} presents the CO (2$-$1) line spectra compared to CO (6$-$5) and [\ion{C}{2}] emission lines of the three targets discussed in this paper. The line widths and redshifts are consistent with each other.  

The CO (2$-$1) line fluxes of the three sources allow us to estimate molecular gas masses. Following \citet{Solomon1992} and assuming a conversion factor $\alpha_{\rm CO}\sim0.8 \ M_{\odot} \ (\rm K\ km\ { s }^{ -1 }\ { pc }^{ 2 })^{-1} $ and $L^{' }_{\rm CO (2-1)} \approx L^{' }_{\rm CO (1-0)}$ \citep{Carilli2013}, we estimate the CO luminosities to be  (21.3 $\pm$ 0.2) $\times$ 10$^{6}$ $L_{\odot}$, (4.0 $\pm$ 0.6) $\times$ 10$^{6}$ $L_{\odot}$ and (6.6 $\pm$ 0.9) $\times$ 10$^{6}$ $L_{\odot}$  and the gas masses to be  (4.3 $\pm$ 0.4) $\times$ 10$^{10}$ $M_{\odot}$, (0.8 $\pm$ 0.1) $\times$ 10$^{10}$ $M_{\odot}$ and (1.3 $\pm$ 0.2) $\times$ 10$^{10}$ $M_{\odot}$ for J2310$+$1855, J0129$-$0035, and J2054$-$0005, respectively. Considering the calibration errors, the gas masses are consistent with those derived from the CO (6$-$5) line emission in \citet{Wang2011} (Table \ref{comeas}), where they assumed  $L^{' }_{\rm CO (6-5)}/L^{' }_{\rm CO (1-0)} = 0.784$ from a Large Velocity Gradient model of J1148+5251 at $z=6.42$ and $\alpha_{\rm CO}\sim0.8 \ M_{\odot} \ (\rm K\ km\ { s }^{ -1 }\ { pc }^{ 2 })^{-1} $.

\subsection{Dust temperature, dust mass and star formation rate of J2310$+$1855}
\label{2310dis}

In this section, we perform a SED fit of J2310$+$1855, for which we have photometry in multiple bands from both this paper and archival data which are all listed in Table \ref{conpho}. We should declare first that our measurements except those from ALMA are all from the unresolved images. We can assume that those measurements represent all the emission from the quasar-galaxy system. However, the ALMA measurements are from marginally or fully resolved data, which may resolve out the low surface density region and result in flux loss. By comparing two flux density measurements in two similar bands - ALMA 263 GHz and IRAM 250 GHz, we find consistent values, which may indicate that almost all the dust emission comes from the central part of the quasar host galaxy. In this case, we may ignore the possible flux loss here. We use the $emcee$ package \citep{emcee2013} to do SED fit to the emission from both star formation and the central AGN based on the method in \citet{Leipski2013, Leipski2014}. In our fitting procedure, we follow \citet{Shangguan2018} to create a likelihood function for the flux upper limits.

We utilize a power law to present the UV/optical emission from the accretion disk and a Clumpy AGN Tori in a 3D geometry (CAT 3D) model \citep{Honig2017} to represent the near-infrared (NIR) and middle-infrared (MIR) contributions from the AGN dust torus, which considers different sublimation radii of various particles. We use two CAT 3D models, one without and one with a polar wind. Unlike  \citet{Leipski2013, Leipski2014}, we do not include a NIR hot black body component from the inner region of the AGN dust torus, but rather an interpolated CAT 3D model described above. We adopt two scenarios for the FIR  dust continuum emission. The first scenario uses a modified black body (MBB) with fixed emissivity index $\beta$ of 1.6 which is a typical value from the high-redshift quasar sample of \citet{Beelen2006}.  The second scenario uses the \citet{Draine2007} dust model which is a linear contribution of PDR and diffuse component models that is described in Equation \ref{dl}: 
\begin{equation}
\label{dl}
\nu F_{\nu, \rm model} = \frac{M_{\rm dust}}{4\pi D_{L}^{2}}[(1-\gamma) \nu p_{\nu}^{(0)}(q_{\rm PAH}, U_{\rm min})+\gamma \nu p_{\nu}(q_{\rm PAH}, U_{\rm min}, U_{\rm max}, \alpha)]
\end{equation}
$q_{\rm PAH}$ is the polycyclic aromatic hydrocarbon (PAH) mass fraction, $U_{\rm min}$ and $U_{\rm max}$ present the minimum and maximum starlight intensity relative to the local interstellar radiation field, $\alpha$ is the power law index of the starlight distribution, $\gamma$ is the dust fraction heated by starlight with $U_{\rm max}>U>U_{\rm min}$, $\nu p_{\nu}^{(0)}$ is the power produced by a single starlight intensity of $U_{\rm min}$ (diffuse component), and $\nu p_{\nu}$ is the power heated by starlight with intensity in the range from $U_{\rm min}$ to $U_{\rm max}$ (PDR component). The parameter details are described in \citet{Draine2007}. As we lack data for the NIR range in which the PAH features appear, we cannot fully constrain the \citet{Draine2007} dust model. The results of our fits are given in Table \ref{sedfitsmea}.

\begin{figure}[htb]
\centering
\includegraphics[scale=0.6]{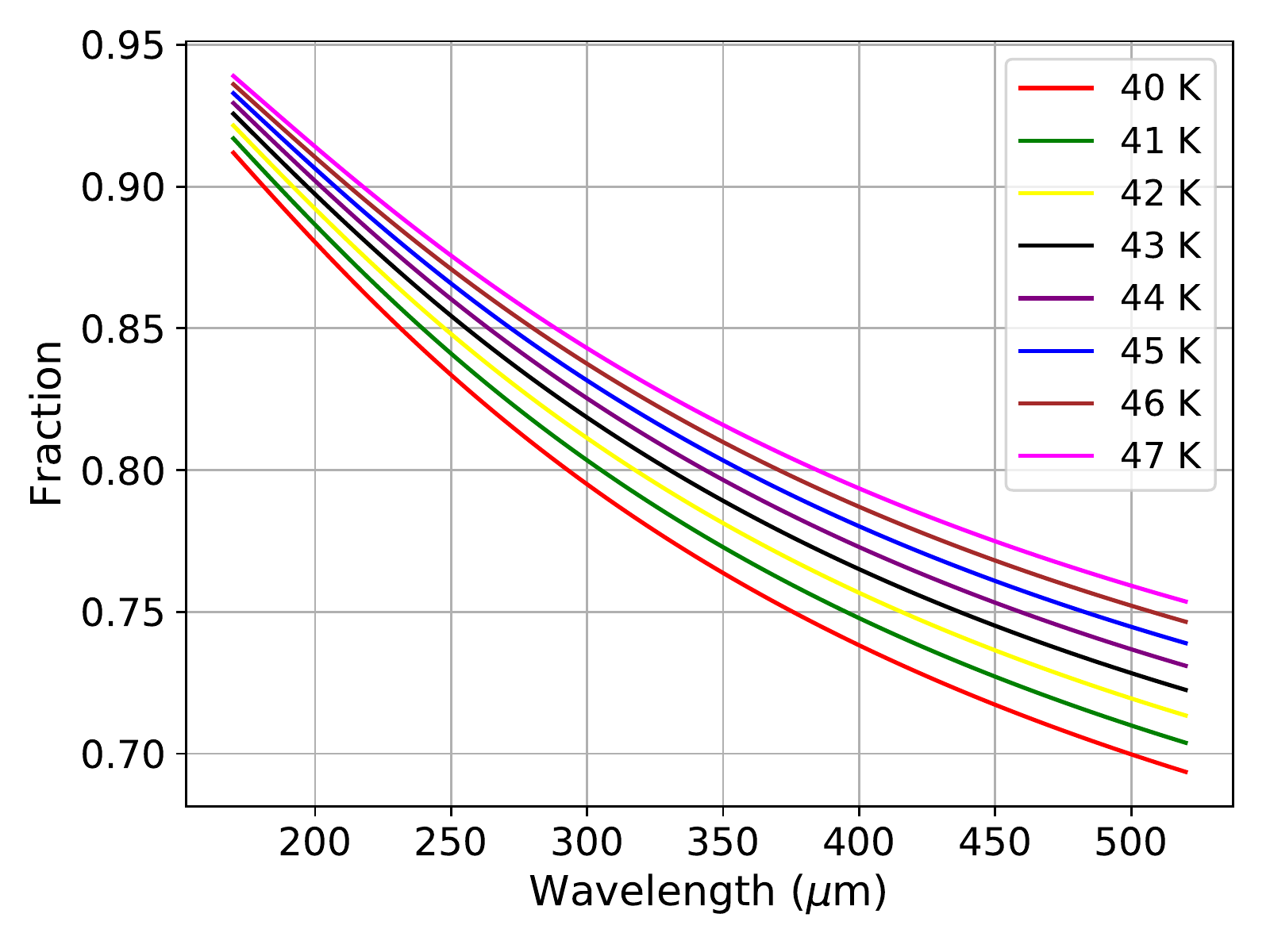}
\caption{The ratio of observed to intrinsic flux of a dust-emitting source considering CMB subtraction (Equation \ref{cmb}) as a function of the dust temperature and wavelength at redshift 6.00. Different colored lines represent different dust temperatures in the range from 40 K to 47 K.}
\label{testcmb}
\end{figure}

As shown in \citet{daCunha2013}, the radiation of the cosmic microwave background (CMB) has two potential effects on our results. One is the heating on the galaxy dust which is important when the CMB temperature is comparable to the dust temperature. At $z$ = 6, the CMB temperature is 2.73 $\times$ (1 + $z$) =  19.11 K, well below our derived temperature of 40 K. So we can neglect the CMB heating. The other is an extra CMB background which will reduce the observed flux density where it is subtracted. The CMB as a background diminishes the observed flux by a fraction defined in Equation \ref{cmb}:
\begin{equation}
\label{cmb}
{\rm Fraction} = \frac{F_{\nu, \rm obs}}{F_{\nu, \rm intrinsic}} = 1 - \frac{B_{\nu}[\lambda, \ T_{\rm CMB}(z)]}{B_{\nu}[\lambda, \ T_{\rm dust}(z)]}
\end{equation}
$F_{\nu, \rm obs}$ and $F_{\nu, \rm intrinsic}$ are the observed and the intrinsic flux densities. $B_{\nu}[\lambda, \ T_{\rm CMB}(z)]$ and $B_{\nu}[\lambda, \ T_{\rm dust}(z)]$ are Planck functions with CMB and galaxy dust temperatures at a given wavelength.  Figure \ref{testcmb} shows this effect at $z = 6$. This figure shows that the longer the wavelength and the cooler the dust temperature, the higher the CMB influence is as a background.  In our analysis, we correct for this effect and get a dust temperature $\sim40$ K for the first scenario of the FIR component. And for the second scenario, we first correct the CMB effect as a background assuming the dust temperature of 40 K which is a typical value from the first scenario and then do the fit. We also estimate the dust temperature based on the average starlight intensity with Equation \ref{dldt1} and \ref{dldt2} and assuming the same $\beta$ = 1.6  with the first scenario in order to make a comparison. 
\begin{equation}
\label{dldt1}
U_{\rm ave} = (1 - \gamma) \times U_{\rm min} + \frac{\gamma \times \ln(\frac{U_{\rm max}}{U_{\rm min}})}{\frac{1}{U_{\rm min}} - \frac{1}{U_{\rm max}}}
\end{equation}
\begin{equation}
\label{dldt2}
T_{\rm DL07} \approx 17.0 \times U_{\rm ave}^{\frac{1}{4 + \beta}} \ \rm K \ ({\rm grain \ size} \ > \ 0.03 \ \mu {\rm m})
\end{equation}
The two scenarios both give a consistent dust temperature $\sim40$ K.

The dust temperature ($\sim40$ K) of J2310+1855 is at the low end of that found for quasars at $z$ $\geq$ 5 (\citealt{Leipski2013, Leipski2014}). We calculate the FIR luminosity by integrating the FIR dust model (the MBB dust model or the \citet{Draine2007} dust model) from 8 to 1000 $\mu$m, and the results are shown in Table \ref{sedfitsmea}. The values from the MBB dust model are (1.4$-$1.6) $\times$ 10$^{13}$ $L_{\odot}$, which would be smaller by a factor of 1.7 if we use a Chabrier initial mass function (IMF; \citealt{Chabrier2003}). Assuming a Salpeter IMF and adopting Equation 4 in \citet{Kennicutt1998}, we derive SFRs of 2400$-$2700 $M_{\odot}$ yr$^{-1}$. The FIR luminosities from the \citet{Draine2007} dust model are (3.2$-$3.6) $\times$ 10$^{13}$ $L_{\odot}$, which are about two times higher than those from the MBB dust model. This may be due to the lack of data in the NIR and MIR parts, where we only have one WISE $w_{3}$ data point which is a 2.5$\sigma$ marginal detection and two PACS data points. These data can poorly constrain the CAT 3D dust torus model as well as the \citet{Draine2007} dust model which contains some PAH lines. More NIR and MIR data would allow us to distinguish these models.

We estimate dust masses (1.7 and 1.6 $\times$ 10$^{9} M_{\odot}$ without and with a polar wind of a CAT 3D torus model) for the first scenario by Equation \ref{eq1}: 
\begin{equation}
\label{eq1}
M_{{\rm dust}} = \frac { F_{125 \mu m}D_{L}^{2} }{ \kappa_{\rm 125 \mu m}B_{\nu}(125 \mu m, T_{{\rm FIR}}) } 
\end{equation}
where $F_{\rm 125 \mu m}$ is the flux density at 125 $\mu$m in the rest frame, $D_{L}$ is the luminosity distance, $\kappa$$_{\rm 125 \mu m}$ = 18.75 cm$^{2}$ g$^{-1}$ is the dust absorption coefficient at 125 $\mu$m \citep{Hildebrand1983}, and $B_{\nu}$ (\rm 125 $\mu$m, $T_{\rm FIR}$) is the Planck value at 125 $\mu$m with  FIR MBB temperature $T_{\rm FIR}$. In the second scenario, the dust masses are directly determined by the fitting process, giving 3.8 and 3.9 $\times$ 10$^{9} M_{\odot}$, respectively. The diffuse component contributes more than $80\%$ of the dust mass and  dominates the luminosity at longer wavelengths $>$100 $\mu$m, however the PDR component dominates the luminosity at shorter wavelengths $<$60 $\mu$m.

The gas-to-dust ratios (GDRs) are $26\pm6$ with dust masses from the first scenario and $11\pm1$ with dust masses from the second scenario. All these ratios are smaller than the widely adopted value - 100$-$150 for the Milky Way, which considers both warm ($>30$K) and cold dust and hydrogen in ionized, atomic and molecular phases. However, GDR values vary in a wide range in different galaxies. For example, \citet{Sandstrom2013} studied 26 nearby star-forming galaxies and proposed a ratio of 91.2. \citet{Devereux1990} calculated a gas to warm dust ratio of 1080 $\pm$ 70 for 58 spiral galaxies. \citet{Baes2014} derived a gas to cold dust ratio $<14.5$ for an early-type galaxy - NGC 5485. \citet{Magdis2011} calculated a ratio $\sim75(35)$ assuming a metallicity of $Z$ = 8.8(9.2) of a starburst galaxy GN20 at $z$ = 4.05 by fitting to the local GDR$-Z$ relation. The large uncertainties in GDRs can arise from determinations of both the gas mass and dust mass. The gas is in multiple phases. A gas mass conversion factor $\alpha$ that just includes molecular gas results in an underestimation and the value of $\alpha$ is likely to be different for different types of galaxies. The dust mass is generally calculated from SED modeling, where small uncertainties in the dust temperature can give large errors. Based on Equation \ref{eq1}, if we increase the $T_{\rm FIR}$ from 40 K to 50 K, the dust mass will decrease by a factor of 2.

\subsection{The Characteristics of the ISM}
\label{dis.3}

\citet{Kaufman1999, Kaufman2006} presented a series of PDR models by solving for the radiation transfer, chemistry equilibrium and thermal balance in a PDR layer. Each model can be described by a constant hydrogen nucleus density $n$ in units of cm$^{-3}$, and the incident FUV intensity $G_{0}$ in units of the Habing field (= 1.6  $\times$  10$^{-3}$ erg cm$^{-2}$ s$^{-1}$; the local Galactic interstellar FUV field). These models can be recreated using the PDR Toolbox (PDRT\footnote{\url{http://dustem.astro.umd.edu/pdrt/}}).  In PDRT modeling, the FIR luminosity is integrated from 30 $\mu$m to infinity. For J2310$+$1855, we estimate the FIR luminosity based on the SED fit result. However, for the remaining two sources, we don't have detailed SED information, so we assume a MBB with a dust temperature of 47 K (the average value of the high redshift quasar sample in \citealt{Beelen2006}), and an emissivity index of 1.6. And we also calculate the FIR luminosities of these two targets with a dust temperature of 40 K from our SED fit of J2310+1855.  As recommended by \citet{Kaufman1999}, we multiply the measured CO line flux by a factor of 2 considering line luminosity from both sides of each cloud (optically thick) when adopting these PDR models.

\begin{figure}
\centering
\includegraphics[scale=0.6]{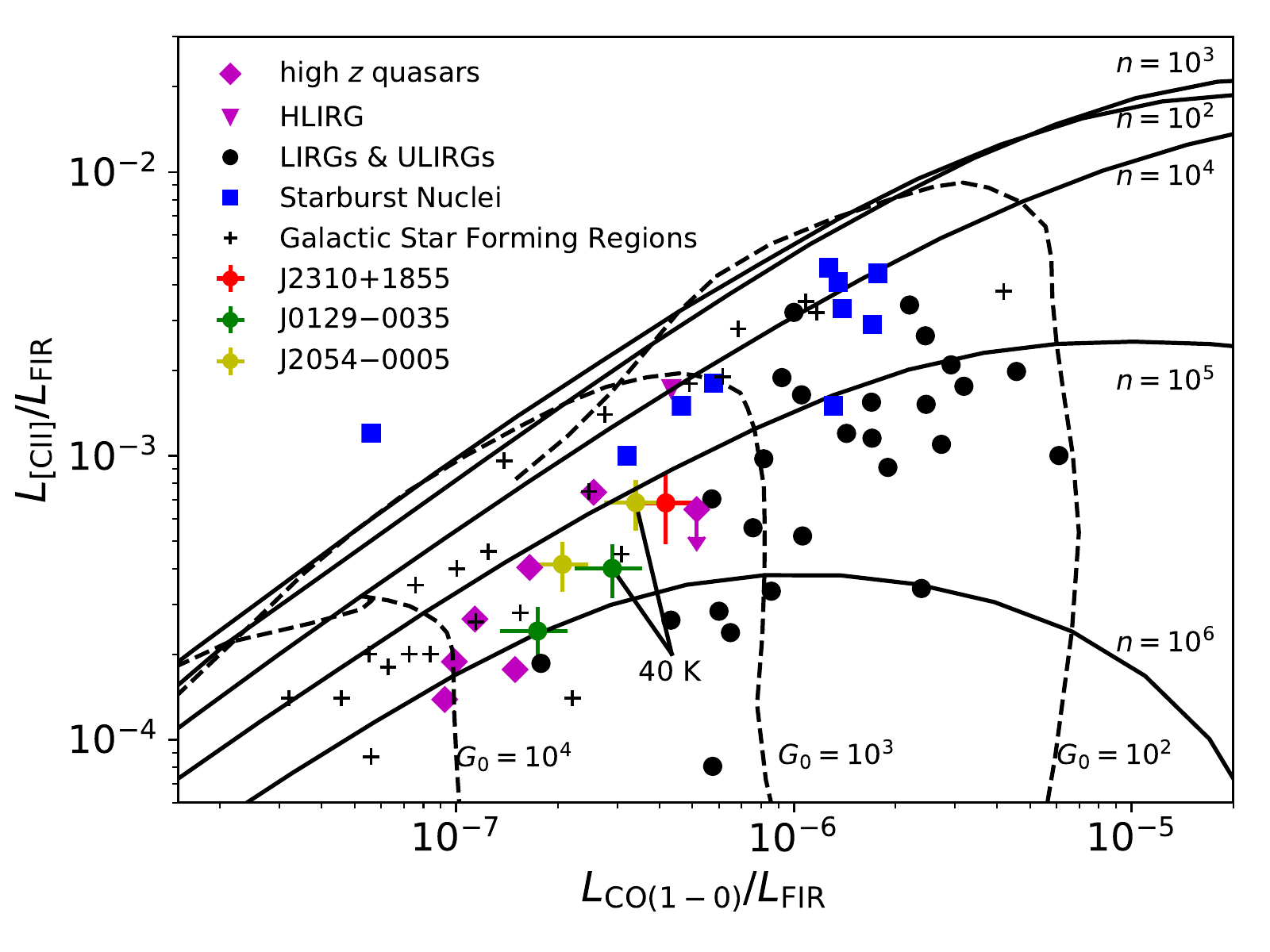}
\caption{$L_{\textsc{\cii\ }}$/$L_{\rm FIR}$ as a function of $L_{\rm CO (1-0)}$/$L_{\rm FIR}$. The curves are PDR models from \citet{Kaufman1999, Kaufman2006}. The black dots are LIRGs and ULIRGs from \citet{Rosenberg2015}. The blue squares represent starburst nuclei in \citet{Stacey1991}. The black crosses present galactic star forming regions \citep{Stacey1991}. The magenta diamonds are high redshift quasars (\citealt{Benford1999}; \citealt{Maiolino2005}; \citealt{Iono2006}; \citealt{Wagg2012, Wagg2014}; \citealt{Leipski2013}; \citealt{Stefan2015}; \citealt{Venemans2017b}; \citealt{WangRan2016}). The magenta triangle is a HLIRG (\citealt{Borys2006}; \citealt{HaileyDunsheath2010}). The red, green and yellow dots with error bars are the three quasars discussed in this paper. In the case of J0129$-$0035 and J2054$-$0005, we also plot them with FIR luminosities assuming a dust temperature of 40 K. To properly compare with the underlying PDR model, we  multiply the CO (1$-$0) line luminosities for all of the plotted samples by a factor of two (see text).} 
\label{ratiomap}
\end{figure}

Figure \ref{ratiomap} shows $L_{\textsc{\cii\ }}$/$L_{\rm FIR}$ as a function of $L_{\rm CO (1-0)}$/$L_{\rm FIR}$ for our three targets and the PDR model grid from \citet{Kaufman1999, Kaufman2006}. We get $L_{\rm CO (1-0)}$ based on an assumption of $L^{\arcmin}_{\rm CO (2-1)}\approx L^{\arcmin}_{\rm CO (1-0)}$ \citep{Carilli2013}.  In this figure, luminous infrared galaxies (LIRGs; $L_{\rm FIR} > 10^{11} L_{\odot}$) and ULIRGs, starburst nuclei and Galactic star forming regions are also plotted.  Our targets and other high redshift quasars at $z > 4$ in the literature (\citealt{Maiolino2005}; \citealt{Iono2006}; \citealt{Wagg2012, Wagg2014}; \citealt{Leipski2013}; \citealt{Stefan2015}; \citealt{WangRan2016}; \citealt{Venemans2017b}) have FIR luminosities ranging from 2.3 $\times$ 10$^{12}$ $L_{\odot}$ to 4.1 $\times$ 10$^{13}$  $L_{\odot}$ and bolometric luminosities in a wide range from 5.7 $\times$ 10$^{12}$ $L_{\odot}$ to 4.3  $\times$ 10$^{14}$  $L_{\odot}$ (\citealt{Priddey2001}; \citealt{Willott2003}; \citealt{Mortlock2011}; \citealt{Wang2013}; \citealt{Wu2015}). These high redshift quasars fall in the same part of the diagram as do the galactic star forming regions with mean and median observed $L_{\textsc{\cii\ }}$/$L_{\rm CO (1-0)}$ values of $\sim3700$ and $\sim3500$, respectively ($\sim4100$ is an empirical value for the local starburst galaxies). In comparing with the PDR models, we constrain the hydrogen nucleus density and the FUV intensity to be a few to ten times 10$^{5}$ cm$^{-3}$ and a few times 10$^{3}$ for our sample.

The flux ratios between CO (6$-$5) and CO (2$-$1) are $8.44\pm1.18$, $10.28\pm2.41$ and $5.67\pm1.50$ for J2310+1855, J0129$-$0035, and J2054$-$0005, respectively. They are similar to that ($= 7.08\pm0.99$) in the well studied quasar J1148+5251 at $z=6.42$. This may indicate that the CO excitation is similar in these quasars at $z\sim6$. With the objects in this paper, there are now eight quasars at $z\sim6$ detected in CO (2$-$1) line emission, and seven of them have CO (6$-$5) detections (\citealt{Bertoldi2003hco}; \citealt{Carilli2007}; \citealt{Wang2010, Wang2011, Wang2011fir, WangRan2016}; \citealt{Stefan2015}). 
The mean and median values of the flux ratio between CO (6$-$5) and CO (2$-$1) of these quasars are $7.85\pm0.98$ and 8.42, respectively. These values are larger than that ($= 4.41\pm1.48$) from the central starburst disk in M82 \citep{Weiss2005} and the mean values ($2.1\pm1.3$ and $3.8\pm0.7$) from a (U)LIRG sample in \citet{Rosenberg2015} and a SMG sample in \citet{Bothwell2013}, which shows that these $z\sim6$ quasars are more highly excited starburst associated systems or that a central AGN may contribute additional heating. A detailed SLED analysis towards J2310+1855 including more high-$J$ CO lines is presented in Li et al. (in preparation).

\section{Summary}
\label{sec5}
We have reported VLA observations of the CO (2$-$1) line emission in three FIR-luminous $z\sim6$ quasars. 
One target (J2310+1855) is marginally resolved at an angular resolution of $\sim0\farcs6$, and the other two (J0129$-$0035 and J2054$-$0005) are point sources at $\sim0\farcs6$ and $\sim2\farcs1$ resolutions respectively. 
CO (2$-$1) line emission is critical to trace the cool gas and to estimate the molecular gas mass directly. 
We have increased the number of CO (2$-$1) detected quasars from 5 to 8 at $z\sim6$. 
The flux ratios between CO (6$-$5) and CO (2$-$1) of the three targets are consistent with that ($\sim7$) of J1148+5251 at $z=6.42$. This may indicate that the CO excitation is similar in these $z\sim6$ quasar host galaxies.
The gas masses based on CO (2$-$1) of the three targets with typical masses of (1$-$4) $\times$ $10^{10} M_{\odot}$ are consistent with those derived by CO (6$-$5). 
Different ISM tracers (CO (2$-$1), CO (6$-$5), and [\ion{C}{2}]) show similar line widths and redshifts of the three quasars. PDR analysis yields very intense FUV fields in all three sources, and indicates similar ISM properties as found in local starburst galaxies. 
We also presented a detailed analysis of the dust continuum of J2310+1855 based on Herschel and ALMA data. 
An SED fit yields a FIR dust temperature of  $\sim40$ K and a SFR of $\sim2400$$-$2700 $M_{\odot}$ yr$^{-1}$, and the derived dust mass is roughly 10$^{9}$ $M_{\odot}$. This is consistent with the strong star formation activity that has been seen in other quasar hosts at $z > 5.7$. The IR data and an SED decomposition are critical to separate the emission from both the central AGN and the star formation activity in the host galaxy.

\acknowledgments

We acknowledge support from the National Key R$\&$D Program of China (2016YFA0400703). R.W. acknowledges supports from the National Science Foundation of China (NSFC) grants No. 11473004, 11533001,  and the National Key Program for Science and Technology Research and Development (grant 2016YFA0400703). D.R. acknowledges support from the National Science Foundation under grant number AST-1614213. This work makes use of the ALMA data: ADS$/$JAO.ALMA$\#$2015.1.01265.S. ALMA is a partnership of ESO (representing its member states), NSF (USA) and NINS (Japan), together with NRC (Canada), MOST and ASIAA (Taiwan), and KASI (Republic of Korea), in cooperation with the Republic of Chile. The Joint ALMA Observatory is operated by ESO, AUI$/$NRAO and NAOJ. The National Radio Astronomy Observatory is a facility of the National Science Foundation operated under cooperative agreement by Associated Universities, Inc.

{\it Facilities:} \facility{Herschel}, \facility{VLA}, \facility{ALMA}.

\bibliographystyle{apj} 
\bibliography{mybib} 

\end{document}